\documentclass[a4paper,11pt]{article}
\pdfoutput=1
 \usepackage{jheppub} 
 \usepackage{graphicx}
\usepackage{amsmath}
 \usepackage{amssymb}
\usepackage{bbold}
 \usepackage{float}
 \usepackage{slashed}
\usepackage{dcolumn}
\usepackage{bm}
\usepackage{color}
\usepackage{multirow}

\allowdisplaybreaks

 \newcommand{\lsim}{{\;\raise0.3ex\hbox{$<$\kern-0.75em\raise-1.1ex\hbox{$\sim$}}\;}}
\newcommand{\gsim}{{\;\raise0.3ex\hbox{$>$\kern-0.75em\raise-1.1ex\hbox{$\sim$}}\;}}
\def\bea{\begin{eqnarray}}
\def\eea{\end{eqnarray}}
\def\bec{\begin{center}}
\def\ec{\end{center}}

\def\beq{\begin{equation}}
\def\eeq{\end{equation}}

\def\bea{\begin{eqnarray}}
\def\eea{\end{eqnarray}}
\def\beq#1\eeq{\begin{align}#1\end{align}}
\def\beqnn#1\eeq{\begin{align*}#1\end{align*}}
\def\ba{\begin{array}}
\def\ea{\end{array}}
\def\bc{\begin{center}}
\def\ec{\end{center}}

\newcommand{\dis}[1]{\begin{equation}\begin{split}#1\end{split}\end{equation}}
\def\vpq{f_{a}}

\newcommand{\C}{\mathbf c}
\newcommand{\Y}{\tilde{\mathbf  y}}
\newcommand{\y}{\mathbf y}
\newcommand{\tr}{\rm tr}

\preprint{CTPU-PTC-21-26}

\title{Precision axion physics with running axion couplings}
 
 \author{
    Kiwoon Choi\footnote{Electronic address: kchoi@ibs.re.kr}, Sang Hui Im\footnote{Electronic address: imsanghui@ibs.re.kr}, Hee Jung Kim\footnote{Electronic address: heejungkim@ibs.re.kr}, and Hyeonseok Seong\footnote{Electronic address: hseong@ibs.re.kr}
 }
     
\affiliation{
 Center for Theoretical Physics of the Universe,  Institute for Basic Science, Daejeon 34126,\\
  South Korea 
    }

\abstract{
We study  the renormalization group running of axion couplings while taking into account  that the Standard Model can be extended to its supersymmetric extension at a certain energy scale below the axion decay constant.
We  then apply our results to three different classes of axion models, i.e.~KSVZ-like, DFSZ-like, and string-theoretic axions, and examine if string-theoretic axions can be distinguished  from others  by having a different pattern of low energy couplings to the photon, nucleons and electron.
We find that the low energy couplings of string-theoretic axions have a similar pattern as those of KSVZ-like axions but yet reveal a sizable difference {which might be testable in future axion search experiments.}  
We also note that the  coupling of KSVZ-like QCD axions to the electron is dominated by a three-loop contribution involving the exotic heavy quark, gluons, top quark and Higgs field. 
}


\vspace{4cm}

\begin{document} 
\maketitle
\flushbottom
 
\section{Introduction}

Axions are one of the most compelling candidates for physics beyond the Standard Model (SM) of particle physics \cite{0807.3125, 1510.07633, 2003.01100}. {Axions} were originally postulated to solve the strong CP problem \cite{Peccei:1977hh, Weinberg:1977ma, Wilczek:1977pj}  and later turned out to be
a natural candidate for dark matter \cite{Preskill:1982cy, Abbott:1982af, Dine:1982ah}.
{Generically there can also be  axions not necessarily} associated with the strong CP problem, which are often dubbed axion-like particle (ALP). Such ALPs can be much lighter than the QCD axion solving the strong CP problem and may constitute dark matter with interesting astrophysical signatures \cite{2101.11735}.

 A key feature of axions is that the field variable is periodic as 
\bea
a(x)\,\cong \, a(x) +2\pi f_a, 
\eea
where the scale parameter $f_a$ is termed the axion decay constant.
Axions can be naturally light as their couplings are constrained by 
an approximate non-linear 
$U(1)_{\rm PQ}$ symmetry  which involves a shift of the axion field:
\bea
U(1)_{\rm PQ}: \quad  a(x) \,\,\, \rightarrow \,\,\, a(x) + {\rm constant}. 
\eea

Axions may originate from the phase of complex scalar field \cite{Peccei:1977hh, Weinberg:1977ma, Wilczek:1977pj,
Kim:1979if, Shifman:1979if, Dine:1981rt, Zhitnitsky:1980tq} as
\bea
\label{field_axion}
\sigma(x) = \rho(x) e^{i a(x)/f_a}, 
\eea
and then $f_a$ is determined by the vacuum expectation value of $\rho$ as $f_a=\sqrt{2}\langle \rho\rangle$.
In the following discussion, such axions will be termed 
 ``field-theoretic axions''. 
It is obvious that the field region around $f_a=0$ for field-theoretic axions is well described by  an effective field theory in which $U(1)_{\rm PQ}$ is linearly realized.
There are two types of field-theoretic axion models which are phenomenologically viable and have been widely discussed in the literature\footnote{
 Here we consider only relatively simple models for field-theoretic axions. 
 For a discussion of more involved models, e.g. those with hierarchical axion couplings,
 see for instance  \cite{2012.05029}. }:
 the Kim-Shifman-Vainshtein-Zakharov (KSVZ) axion models \cite{Kim:1979if, Shifman:1979if} in which all SM fields are neutral under the linearly-realized $U(1)_{\rm PQ}$ 
and the Dine-Fischler-Srednicki-Zhitnitsky (DFSZ) axion models \cite{Dine:1981rt, Zhitnitsky:1980tq} in which 
 the SM fermions carry nonzero $U(1)_{\rm PQ}$-charges.

Axions can also arise from the zero modes of a higher-dimensional $p$-form gauge field $C^{(p)}$
which 
has a quantized coupling to $(p-1)$-brane in the underlying UV theory \cite{hep-th/0605206, 0905.4720}: 
\bea
\label{string_axion}
a(x)=\int_{\Sigma_p} C^{(p)}_{[m_1,..,m_p]}(x,y)\,dy^{m_1}...dy^{m_p},
\eea
where $\Sigma_p$ is a $p$-dimensional cycle in the compact internal space with the coordinates $y^m$.
We will term such axions originated from $C^{(p)}$ 
``string-theoretic axions''. It is widely believed that the limit with
 $f_a=0$ for string-theoretic axions is in swampland at an infinite field distance with infinite towers of light particles \cite{1903.06239}.

 It is well known that KSVZ axions and DFSZ axions have a different pattern of low energy couplings,
 {which might be testable in future axion search experiments}
 \cite{0807.3125, 2003.01100}. For instance,  at tree level,  the couplings of KSVZ axions to the SM fermions are all vanishing, while
those of DFSZ axions are of order unity in the unit of $1/f_a$, which results in different ratios among  the low energy axion couplings $g_{aX}$ ($X=\gamma, p,n,e$) to the photon, nucleons and electron.
Similarly, generic DFSZ-like axions with ${\cal O}(1)$ tree-level couplings (again in the unit of $1/f_a$) to the light quarks and electron have a clearly distinguishable pattern of $g_{aX}$
($X=\gamma, p,n,e$) compared to those of generic KSVZ-like axions with negligible tree-level couplings to the light quarks and electron. 
 On the other hand, it has not been examined yet if  string-theoretic axions can also be distinguished from other types of axions by their low energy couplings. 
 As we will see, the couplings of string-theoretic axions to the SM gauge and matter fields have a similar
 pattern as those of KSVZ-like axions. Therefore it requires a careful analysis  
 including radiative corrections \cite{Srednicki:1985xd, hep-ph/9306216, 1708.00021, 2002.04623, 2011.08205, 2012.05029, 2012.09017, 2012.12272} to axion couplings  to see if string-theoretic axions can be discriminated from KSVZ-like axions.

With this motivation,  in this paper, we study
 the renormalization group (RG) running of axion couplings in a generic context and apply the results to
 the models of our interest.
To examine if  string-theoretic axions  and KSVZ-like axions
 can have a distinguishable pattern of  $g_{aX}$ ($X=\gamma, p, n, e$),
we first perform a generic renormalization group analysis for the axion couplings to matter fermions and Higgs fields
 while taking into account that the Standard Model can be extended to its supersymmetric extension at a scale below $f_a$.   
We then find that  string-theoretic QCD axion and KSVZ-like QCD axion have a significantly different value of the coupling ratio $g_{ae}/g_{a\gamma}$, which would make it feasible to distinguish one from the other by an experimental measurement of  $g_{ae}/g_{a\gamma}$.
 For an axion-like particle (ALP) which does not couple to the gluons but has a non-zero coupling to the photon, the prospect for discrimination is much more promising: {all of the three coupling ratios $g_{aY}/g_{a\gamma}$ ($Y=p,n,e$) for
  string-theoretic ALP and KSVZ-like ALP differ by about one order of magnitude.}  We also find that the coupling of KSVZ-like QCD axion to the electron is dominated
 by a three-loop contribution involving the heavy exotic quark, gluons, top quarks and Higgs doublet, which was not noticed in the previous studies \cite{Srednicki:1985xd, hep-ph/9306216, 1708.00021, 2002.04623, 2011.08205, 2012.09017, 2012.12272}.

 The organization of this paper is as follows. In section \ref{sec:UV}, we discuss the relevant features of the axion couplings to gauge and matter fields
 at the UV boundary scale $\mu\sim f_a$ for {DFSZ-like, KSVZ-like} and string-theoretic axions. In section \ref{sec:RG}, we present the RG equations for axion couplings in the context of generic effective lagrangian and compute the resulting radiative corrections to the low energy axion couplings to the light quarks and electron.
 In this analysis,
 we take into account  that the SM can be extended to its supersymmetric extension at an energy scale below $f_a$. 
 In section \ref{sec:PP}, we apply our result to {the three different classes of axions, i.e. DFSZ-like, KSVZ-like, and string-theoretic axions,} to see if they can be distinguished from each other by their low energy couplings to the photon, nucleons and  electron. Section \ref{sec:CC} is the conclusion.

\section{Axion couplings to matter and gauge fields at UV scales} \label{sec:UV}

In this section, we discuss  the axion couplings to gauge and matter fields at the UV boundary scale $\mu\sim f_a$ in models of 
{KSVZ-like, DFSZ-like}, and string-theoretic axions.
We will be brief on field-theoretic {KSVZ-like and DFSZ-like} axions and pay more attention to the characteristic features of string-theoretic axions.
To proceed, we adopt the Georgi-Kaplan-Randall (GKR) field basis \cite{Georgi:1986df} which can be defined at a scale
around or below $f_a$,  for which {\it only}
the axion field transforms under 
$U(1)_{\rm PQ}$ as
\bea
\label{gkr_pq}
U(1)_{\rm PQ}:\,\,\,\, a(x)\rightarrow a(x)+{\rm constant},
\eea
and all other fields in the model are invariant under $U(1)_{\rm PQ}$.
Then the relevant axion couplings at $\mu\sim f_a$  take the form
  \bea
  \label{axion_coupling}
\frac{1}{2}  \partial_\mu a \partial^\mu a
+\frac{\partial_\mu a}{f_a}\Big[ i c_\phi(\phi^{*} D^\mu \phi-\phi D^\mu\phi^{*}) + c_{\psi}\bar\psi\bar\sigma^\mu \psi\Big] + c_A\frac{g_A^2}{32\pi^2}\frac{a(x)}{f_a} F^{A\mu\nu}\tilde F^A_{\mu\nu},
\eea 
where $\phi$ and $\psi$ denote {canonically normalized} gauge-charged  scalar fields and chiral fermions
in the model, and $F^A_{\mu\nu}=(G^{\alpha}_{\mu\nu}, W^i_{\mu\nu}, B_{\mu\nu})$ are the {canonically normalized} gauge field strength
of the SM gauge group $SU(3)_c\times SU(2)_L\times U(1)_Y$. In this description,
the axion couplings $c_A$ ($A=G,W,B$) to gauge fields are quantized, e.g. integer-valued for
an appropriate normalization of the gauge couplings $g_A^2$, and invariant under the subsequent RG evolution, while
the couplings $c_{\phi}$ and $c_\psi$ to matter fields are real-valued and generically have nontrivial RG evolution induced by the gauge and Yukawa interactions in the model.

As will be discussed in detail in section \ref{sec:RG}, to examine the low energy axion couplings in
 the model,
one can first scale the effective lagrangian 
(\ref{axion_coupling}) down to a low energy scale $\mu={\cal O}(1)$ GeV.  The relevant axion couplings 
at this scale are given by
\bea
\label{coupling_gev}
\frac{1}{32\pi^2}\frac{a(x)}{f_a}\left( c_\gamma e^2 F^{\mu\nu}\tilde F_{\mu\nu} + c_G g_s^2G^{\alpha\mu\nu}\tilde{G}^{\alpha}_{\mu\nu}\right)
 +\sum_{\Psi=u,d,e} \frac{\partial_\mu a}{2f_a} C_\Psi \bar\Psi\gamma^\mu\gamma_5\Psi,\eea
 where {$e$ and $g_s$ are the electromagnetic and color gauge couplings, respectively, and}
 \bea
 c_\gamma =c_W+c_B,\quad
  C_\Psi = C_\Psi^0 +\Delta C_\Psi,\eea
 with
 \bea
 \label{tree_coupling}
 C^0_u&=&c_{Q_1}(f_a)+c_{u^c_1}(f_a)+c_{H}(f_a),\nonumber \\
 C^0_d&=&c_{Q_1}(f_a)+c_{d^c_1}(f_a)-c_{H}(f_a), \nonumber \\
 C^0_e&=&c_{L_1}(f_a)+c_{e^c_1}(f_a)-c_{H}(f_a).
 \eea
 Here $C_\Psi^0$ ($\Psi=u,d,e$) denote the axion couplings to the light quarks and electron evaluated at {\it tree-level}, {which can be interpreted as the couplings at the UV boundary scale $\mu= f_a$ in our approximation}, $\Delta C_\Psi$ are  radiative corrections to $C_\Psi$ received over the scales from $f_a$ to $\mu={\cal O}(1)$ GeV, which will be extensively discussed in section \ref{sec:RG}, {
 $c_\phi(f_a)$ and $c_\psi(f_a)$ are the
 axion couplings 
  at $\mu=f_a$
   including the couplings to
  the Higgs doublet $H$ and the three generations of chiral quarks and leptons  $\psi=\{Q_i, u^c_i, d^c_i, L_i, e^c_i\}$ ($i=1,2,3$), which would be determined by the underlying UV-completed  axion model},  and the effects of flavor mixing are ignored for simplicity.
The axion-Higgs coupling $c_H$ gives rise to the axion couplings to the SM fermions via the axion-$Z$ boson kinetic mixing $\partial_\mu a Z^\mu$ below the weak scale. Alternatively, one can rotate away the axion-Higgs coupling $c_H$ to the axion-fermion couplings by performing axion-dependent field redefinition proportional to $U(1)_Y$ hypercharge:
 \dis{
 H \rightarrow H e^{i c_H a/f_a},\quad \psi \rightarrow \psi  e^{i (Y_\psi/Y_H) c_H a/f_a},
 }
 which results in Eq. (\ref{tree_coupling}).

In regard to the experimental verification of axions, the most relevant couplings are those to the photon, nucleons and electron at scales well below GeV:
 \bea
 \label{observable_coupling}
 \frac{1}{4}g_{a\gamma} a\vec E\cdot \vec B + \partial_\mu a \left[
 \frac{g_{ae}}{2m_e} \bar e \gamma^\mu \gamma_5 e+ \frac{g_{an}}{2m_n} \bar n \gamma^\mu \gamma_5 n+\frac{g_{ap}}{2m_p} \bar p \gamma^\mu \gamma_5 p\right] \eea
 which are determined by the couplings in (\ref{coupling_gev}) as follows:
 \bea
 g_{a\gamma} &\simeq &\frac{\alpha_{\rm em}}{2\pi}\frac{1}{f_a}\Big( c_\gamma -\frac{2}{3}\frac{m_u+4m_d}{m_u+m_d} c_G\Big)\,
 \simeq\,  \frac{\alpha_{\rm em}}{2\pi}\frac{1}{f_a}\Big( c_\gamma-1.92 c_G\Big),\nonumber \\
 g_{ap} &\simeq & \frac{m_p}{f_a}\left( C_u \Delta u + C_d \Delta d -\Big(\frac{m_d}{m_u+m_d}\Delta u +\frac{m_u}{m_u+m_d}\Delta d\Big) c_G\right),\nonumber \\
 &\simeq &  \frac{m_p}{f_a}\Big(  0.90 \,C_u(2\, {\rm GeV}) -0.38\, C_d(2\,{\rm GeV}) -0.48 \,c_G\Big),\nonumber \\
 g_{an} &\simeq&  \frac{m_n}{f_a}\left( C_d \Delta u + C_u \Delta d -\Big(\frac{m_u}{m_u+m_d}\Delta u +\frac{m_d}{m_u+m_d}\Delta d\Big) c_G\right), \nonumber \\
 &\simeq & \frac{m_n}{f_a}\Big(0.90 \,C_d(2\, {\rm GeV})-0.38\, C_u(2\,{\rm GeV})-0.04 \,c_G\Big), \nonumber \\
 g_{ae} &\simeq& \frac{m_e}{f_a} C_e(m_e),\eea
 with $\Delta u = 0.897(27)$ and $\Delta d = -0.376(27)$ at $\mu = 2$ GeV in $\overline{\rm MS}$, 
 and $m_u/m_d = 0.48(3)$   \cite{1511.02867, 2003.01100}.
One of our primary concerns is if string-theoretic axions can be
discriminated from field-theoretic axions by having a distinguishable  pattern of $g_{aX}$ ($X=\gamma,p,n,e$).

\subsection{Field-theoretic axions: KSVZ-like and DFSZ-like axions}

Field-theoretic axion models have a UV completion with a linearly realized Peccei-Quinn symmetry:
\bea
\label{linear_pq}
U(1)_{\rm PQ}:\quad \sigma \rightarrow e^{i\alpha} \sigma, \quad 
\phi\rightarrow e^{iq_\phi\alpha}\phi, \quad \psi\rightarrow e^{iq_\psi\alpha}\psi, \eea
where $\sigma$ denotes a $U(1)_{\rm PQ}$-charged complex scalar field whose phase field is identified as the axion as in (\ref{field_axion}).  
For simplicity, here we assume that $U(1)_{\rm PQ}$ is spontaneously broken predominantly by the vacuum expectation value of a single gauge-singlet complex scalar field $\sigma$. 

To move to the GKR field basis \cite{Georgi:1986df}, one can parameterize $\sigma$ as
\bea
\sigma = \frac{1}{\sqrt{2}}(f_a +\hat\rho) e^{ia/f_a},\eea
where $\langle \hat\rho \rangle =0$,
and make an axion-dependent field redefinition: \bea
\label{field_re}
\phi\rightarrow e^{-iq_\phi a/f_a}\phi, \quad \psi\rightarrow e^{-iq_\psi a/f_a}\psi.
\eea
Then, $U(1)_{\rm PQ}$ is realized as the GKR form in (\ref{gkr_pq}), and 
the axion couplings to gauge and matter fields arise as a consequence of the field redefinition (\ref{field_re}). The resulting couplings  at 
$\mu\sim f_a$ are determined simply by the PQ-charges of the linear $U(1)_{\rm PQ}$ symmetry (\ref{linear_pq}):
\bea
c_{\phi}(f_a)= q_{\phi},\quad c_{\psi}(f_a)= q_{\psi}, \quad  {c_A=\sum_\psi 2q_\psi {\rm Tr}(T_A^2(\psi))},\eea
where $T_A(\psi)$ $(A=G,W,B)$  denotes the gauge charge of $\psi$ for the SM gauge group
$SU(3)_c\times SU(2)_L\times U(1)_Y$.

There are two different types of field-theoretic axion models that have been widely discussed in the literature, 
DFSZ models 
 in which the SM fermions carry non-zero $U(1)_{\rm PQ}$ charges and KSVZ models
in which all SM fermions are neutral under $U(1)_{\rm PQ}$.
{A minimal non-supersymmetric (non-SUSY) DFSZ model \cite{Dine:1981rt, Zhitnitsky:1980tq} 
involves the two Higgs doublets $H_u$ and $H_d$ which have the Yukawa couplings for the up-type quark masses and the down-type quark and lepton masses, respectively, as well as the coupling $\kappa \sigma^2 H_uH_d$ in the potential with $\kappa$ small enough to allow  the correct electroweak symmetry breaking. The resulting
   axion
couplings at $\mu\sim f_a$ are given by
\bea \label{DFSZpar}
\mbox{non-SUSY DFSZ}: \quad c_{H_{u,d}}(f_a)=-1, \quad c_{\psi}(f_a) =\frac12, \quad  {c_G=c_W=\frac{3}{5}c_B=6},
\eea
where  $\psi=
\{Q_i, u^c_i, d^c_i, L_i, e^c_i\}$ are the chiral quarks and leptons in the SM. The corresponding
tree-level axion couplings to the light quarks and electron, which can be identified as  the couplings at $\mu\sim f_a$,
are 
\bea \label{DFSZtree}
\mbox{DFSZ}: \quad 
C_u^0= 2\cos^2 \beta, \quad C_d^0=C_e^0= 2\sin^2 \beta,\eea
where $\tan\beta =\langle H_u\rangle/\langle H_d\rangle$ and we used 
$c_H = c_{H_u}\sin^2\beta -c_{H_d}\cos^2\beta$ for the coupling to
the SM Higgs doublet $H$. It is straightforward to supersymmetrize this model, for instance with a superpotential involving a term {$\sim \sigma^2 H_uH_d/M_P$ \cite{Kim:1983dt}, where
$M_P=2.4\times 10^{18}$ GeV is the reduced Planck scale}. 
The axion couplings of this SUSY DFSZ model at $\mu\sim f_a$ are
  \bea \label{SUSYDFSZpar}
\mbox{SUSY DFSZ}: \quad c_{H_{u,d}}(f_a)=-1, \quad c_{\psi}(f_a) =\frac12, \quad  c_G=6, {\quad c_W=4,\quad c_B=8},
\eea
where now $H_{u,d}$ and $\psi$ 
denote the full supermultiplets including superpartners, and $c_{W,B}$ includes the Higgsino contribution to the $U(1)_{\rm PQ}$ anomaly.
Obviously, the tree-level axion couplings to the light quarks and electron in this SUSY DFSZ 
model are the same as those of its non-SUSY counterpart which is given by
(\ref{DFSZtree}).

In 
KSVZ models \cite{Kim:1979if, Shifman:1979if}, all SM fields are neutral under $U(1)_{\rm PQ}$. However, there exist exotic 
$U(1)_{\rm PQ}$-charged left-handed quark and anti-quark $({\cal Q}, {\cal Q}^c)$ which get a heavy mass  as a  consequence of the spontaneous breakdown of $U(1)_{\rm PQ}$ by the vacuum  expectation value  $\langle \sigma \rangle =f_a/\sqrt{2}$, e.g. get a mass by the Yukawa coupling
\bea \label{KSVZmod}
{\cal L}_{\rm Yukawa}= y_{\cal Q} \sigma {\cal Q}{\cal Q}^c.\eea
One then finds
\bea
{\rm KSVZ}:\quad 
c_{{\cal Q}}(f_a) =c_{{\cal Q}^c}(f_a)=\frac{1}{2}, \quad  c_H(f_a)=c_{\psi}(f_a)=0, \quad  { c_A= 2 {\rm Tr}(T_A^2({\cal Q}))}\eea
for $\psi=
\{Q_i, u^c_i, d^c_i, L_i, e^c_i\}$, which result in 
\bea
{\rm KSVZ}:\quad
C^0_\Psi=0 \quad (\Psi=u,d,e).\eea
In fact, there can be a contribution to $c_{\psi}(f_a)$ from Planck-scale-suppressed higher-dimensional operator such as $\sigma^*\partial_\mu\sigma \bar \psi\sigma^\mu \psi/M_P^2$, which would
give $c_\psi(f_a)= {\cal O}(f_a^2/M_P^2)$, and therefore 
\bea
C_\Psi^0={\cal O}\Big(\frac{f_a^2}{M_P^2}\Big)\lesssim {\cal O}\left(\Big(\frac{g_{\rm GUT}^2}{8\pi^2}\Big)^2\right),\eea 
where $g_{\rm GUT}$ denotes the SM gauge coupling around the GUT scale or $f_a$, and
we use 
 the axion weak gravity conjecture bound $f_a/M_P\lesssim {\cal O}(g_{\rm GUT}^2/8\pi^2)$ \cite{hep-th/0601001} for the last inequality.
As in the case of DFSZ model, it is straightforward to supersymmetrize the above KSVZ model and the resulting SUSY  model also has vanishing tree-level couplings to the SM Higgs and fermion fields.
}

One can consider  {more generic models of field-theoretic axions other than those} presented above, e.g. models with different Higgs sector for the Yukawa couplings of the SM fermions, or  additional $U(1)_{\rm PQ}$-charged fields, or
even flavor-dependent  $U(1)_{\rm PQ}$-charges \cite{1708.02111, 1610.01639, 1610.07593, 1709.07039, 2101.03173, 1612.05492, 1612.08040, 1712.04940}.  In the following, we use
the term ``{DFSZ-like axions}'' for {\it generic} field-theoretic axions with 
$C_\Psi^0={\cal O}(1)$ ($\Psi=u,d,e$) and the term ``{KSVZ-like axions}'' for those with $C_\Psi^0=0$ ({or more generically $C_\Psi^0\lesssim{\cal O}\big((g_{\rm GUT}^2/8\pi^2)^2\big)$}. As is well known (e.g. see Table 1 in \cite{1801.08127}), {DFSZ-like and KSVZ-like axions} have a clearly different pattern of the observable low energy couplings $g_{aX}$ ($X=\gamma, p,n,e$),  so
can be easily distinguished from each other.

\subsection{String-theoretic axions}

String theory involves a variety of extended objects ($(p-1)$-branes) which have quantized couplings to a $p$-form gauge field in the theory \cite{Polchinski:1998rr}. {As a consequence,   string models accommodating the SM generically contain a string-theoretic axion that couples to the SM gauge and matter fields
in 4-dimensional effective theory \cite{hep-th/0605206, 0905.4720}.}
Such axions are accompanied by their scalar partners, i.e. saxions or moduli $\tau$, whose vacuum expectation value can be identified as the Euclidean action of the brane instanton which has a quantized coupling to axion.

{Generically  there can be multiple string-theoretic axions, sometimes many  \cite{0905.4720}.
Yet the low energy consequence of those  axions 
depends crucially on the axion masses which  have a close connection with  the mechanism to stabilize their moduli partners.
For instance, 
many of string-theoretic axions may get a heavy mass together with their moduli partners and therefore
 are decoupled from the low energy world.
On the other hand,
in this paper  we are mostly interested in light axions  whose couplings to the SM can be probed by precision low energy experiments, which include a QCD axion and an ultralight ALP coupled to the photon. Such light axions
are required to be much lighter than their moduli partners, so we need
a mechanism to stabilize moduli while keeping some axions to be nearly massless.  In the following, we simply assume that
underlying string model provides such a mechanism as was discussed for instance in \cite{hep-th/0602233, hep-th/0611279, 1003.1982, 1004.5138, 1206.0819, 1404.3880}.
    }

Usually, string-theoretic axion and its modulus partner $\tau$ form a complex scalar field which corresponds to the scalar component of a chiral superfield in 4-dimensional effective supergravity of the underlying string compactification:
\bea
T=\tau(x) +i\frac{a(x)}{f_a}.
\eea
Then the axion couplings to gauge and matter fields are determined by the moduli Kahler potential $K_0$, the matter Kahler metric $Z_I$,  and the holomorphic gauge kinetic functions ${\cal F}_A$ of the model,
  which are given by
\bea
\label{4d_sugra}
 K&=&K_0(T+T^*) + Z_I(T+T^*)\Phi^{*}_I \Phi_I + ...,\nonumber \\
{\cal F}_A &=& \frac{c_A}{8\pi^2}T + ..., 
\eea 
where  $\Phi_I$ denote gauge-charged chiral matter superfields in the model, and the ellipses  stand for the terms higher order in $\Phi_I$, as well as the non-perturbative $U(1)_{\rm PQ}$-breaking terms which would be exponentially suppressed by $e^{-T}$.
Here, for simplicity, we assume a diagonal form of the  matter Kahler metric $Z_I$  and ignore moduli superfields other than $T$.
Note that the factor $1/8\pi^2$ in ${\cal F_A}$ appears due to the normalization convention of $T$ for which
${\rm Im}(T)=a(x)/f_a\cong {\rm Im}(T) +2\pi$, and $c_A$ are rational numbers (or integers)
in this convention.
It is then straightforward to find that the decay constant and couplings of the axion $a(x)$ are given by 
\bea
\frac{1}{2}f^2_a = M_P^2\frac{\partial^2K_0}{\partial T\partial T^*}\eea
and 
\bea \label{string_bc}
 c_{\phi_I} = \frac{\partial\ln Z_I}{\partial T}, \quad c_{\psi_I} = \frac{\partial \ln(e^{-K_0/2} Z_I)}{\partial T}, \quad c_A = 8\pi^2\frac{\partial}{\partial T} {\cal F}_A,\eea
 where $\phi_I$ and $\psi_I$ denote the scalar component  and the fermion component, respectively, of the chiral superfield $\Phi_I$.

The vacuum value of $\tau$ can be constrained in several ways. As noticed above, $\tau$ corresponds to
the Euclidean action of a brane instanton which couples to $a(x)$, and then
 the axion weak gravity conjecture (WGC) \cite{hep-th/0601001} implies 
\bea
\tau \,\lesssim\, {\cal O}\Big(\frac{M_P}{f_a}\Big).\eea 
Generically this brane instanton 
 contributes to the axion potential, which can be parameterized as  
\bea
\label{brane_instanton}
\delta V = e^{-\tau}\Lambda^2_{\rm ins} M_P^2\cos \Big(\frac{a}{f_a}\Big), \eea
where $\Lambda_{\rm ins}$ is a model-dependent mass parameter determined by a more detailed feature of the 
model, in particular by the fermion zero modes of the brane instanton and the scale of supersymmetry breaking.
In fact, explicit examples in string theory suggest that those brane instantons induce a correction of ${\cal O}(e^{-T})$ to the superpotential or to the Kahler potential of the 4-dimensional effective supergravity \cite{Dine:1986zy, 0902.3251}, which results in\footnote{{For models with a string scale $M_{\rm st}\ll M_P$, there can be an additional suppression \cite{hep-th/0602233} which is not crucial for our subsequent discussion.}} 
\bea
\Lambda_{\rm ins}^2= {\cal O}(m_{3/2}M_P) ~\,\, {\rm or} ~\,\, {\cal O}(m_{3/2}^2).\eea
At any rate, the brane-instanton-induced potential (\ref{brane_instanton}) provides a lower bound on the axion mass:\bea
m_a^2 \gtrsim e^{-\tau}\Lambda_{\rm ins}^2M_P^2/f_a^2,\eea
which implies that
the brane instanton action is bounded from below as 
\bea
\label{lower_bound1}
\tau \,\gtrsim\, 2\ln (\Lambda_{\rm ins}/m_a).\eea
For a string-theoretic QCD axion,
the  brane instanton-induced axion potential (\ref{brane_instanton}) should be further suppressed not to spoil the axion solution  to the strong CP problem. Specifically one needs \bea
\delta V\,\lesssim\, 10^{-10}m_\pi^2 f_\pi^2,
\eea
implying
\bea
\tau \,\gtrsim\, 
 \ln (10^{10} \Lambda_{\rm ins}^2/f_\pi m_\pi)
 \label{lower_bound}\eea
for a QCD axion,
where $m_\pi$ and $f_\pi$ are the pion mass and the pion decay constant, respectively.

On the other hand, for $c_A$ of order unity, $\tau$ is also bounded from above:
\bea
\tau \,\lesssim\, {\cal O}\Big(\frac{8\pi^2}{g^2_{\rm GUT}}\Big)
\label{upper_bound}\eea
in order to avoid a too small unified gauge coupling   
 $g^2_{\rm GUT}\simeq 1/{\rm Re}({\cal F}_A)$.  
For  $\Lambda_{\rm ins}\gtrsim m_{3/2}$, the lower bound (\ref{lower_bound1}) or (\ref{lower_bound}) and the upper bound (\ref{upper_bound}) are numerically comparable to each other. This implies
\bea
\tau ={\cal O}\Big(\frac{8\pi^2}{g_{\rm GUT}^2}\Big)\eea
{at least for light string-theoretic axions which can be identified as a  QCD axion or 
an ALP constituting dark matter.}
 {Then, with such a large value of $\tau$, the following order of magnitude estimates are expected to be held}:
\bea
\label{estimate0}
&&\frac{\partial K_0}{\partial\tau}  \,\sim\, \tau \frac{\partial^2K_0}{\partial\tau^2}\,\sim\,
\tau \frac{f_a^2}{M_P^2}\,\lesssim\,
{\cal O}\Big(\frac{g_{\rm GUT}^2}{8\pi^2}\Big),\nonumber \\
&& \frac{\partial \ln Z_I}{\partial\tau} \,\sim\, \frac{1}{\tau} \,=\, {\cal O}\Big(\frac{g_{\rm GUT}^2}{8\pi^2}\Big),
\eea
which suggest {
\bea
\label{estimate}
c_\phi(f_a)= \omega_\phi \frac{g_{\rm GUT}^2}{16\pi^2}, \quad  c_\psi(f_a) = \omega_\psi\frac{g_{\rm GUT}^2}{16\pi^2},
\eea
with \bea \label{estimate_omega}
\omega_\phi \sim \omega_\psi ={\cal O}(1).\eea
}
Accordingly, the {\it tree-level} couplings $C_\Psi^0$ ($\Psi=u,d,e$) of string-theoretic axions to the light quarks and electron, which are defined in (\ref{tree_coupling}), are estimated as \bea
\label{estimate1}
\mbox{String-theoretic}:\quad C_\Psi^0=\,{\cal O}\Big(\frac{g_{\rm GUT}^2}{16\pi^2}\Big) \quad (\Psi=u,d,e).\eea
{Note that this estimate applies for string-theoretic axions which have a non-zero coupling to at least one of the SM gauge fields  and also  light enough
to be identified as a QCD axion or an ALP dark matter.}

The above estimate indicates that the couplings of string-theoretic axions have a similar pattern 
 as those of {KSVZ-like axions} in the sense that both axions have  $|C_\Psi^0/c_A|\ll 1$.
Yet one might be able to distinguish one from the other 
{\it if} the radiative corrections to $C_\Psi$ 
are significantly smaller
than the tree-level couplings $C_\Psi^0$ for string-theoretic axions. In such case, the low energy couplings $C_\Psi$
of {KSVZ-like axions} will be significantly smaller than those of string-theoretic axions, which would leave a notable imprint 
in the observable axion couplings $g_{aX}$ ($X=\gamma, p,n,e$)
in (\ref{observable_coupling}).
To see if {KSVZ-like axions} and string-theoretic axions indeed have a distinguishable pattern of $g_{aX}$,
we will conduct in section \ref{sec:RG} an RG analysis for the axion couplings to matter fields.

In fact, the  estimate (\ref{estimate}) can be elaborated with the scaling argument made in \cite{hep-th/0609180}. 
For gauge fields propagating over a $d_{\rm YM}$-dimensional cycle $\Sigma_{\rm YM}$ in  6-dimensional internal space\footnote{In heterotic string theory, gauge fields propagate over 
the entire 6-dimensional internal space, so
$d_{\rm YM}=6$. In other string theories with $D$-branes, gauge fields are confined on $D_p$-branes wrapping a $(p-3)$-dimensional cycle, so $d_{\rm YM}=p-3$.}, the corresponding gauge couplings scale as \bea
{\rm Re}({\cal F}_A)=\frac{1}{g_A^2} \propto {\cal V}(\Sigma_{\rm YM})\eea
at leading order  in $1/{\cal V}(\Sigma_{\rm YM})$,
which ignores for instance
the flux densities in $\Sigma_{\rm YM}$, 
where ${\cal V}(\Sigma_{\rm YM})$ is the volume (in the string frame)
of $\Sigma_{\rm YM}$. 
On the other hand, the physical Yukawa couplings
 $y_{IJK}$  are determined as  
\bea
y_{IJK} =\int_{\Sigma_{IJK}}  \phi_I^{(0)}\phi_J^{(0)}\phi_K^{(0)} \, d{\cal V}(\Sigma_{IJK}),\eea
where $\phi_I^{(0)}$ 
is the zero mode wavefunction of $\Phi^I$ which can propagate over  a $d_I$-dimensional subcycle
$\Sigma_I$ 
of $\Sigma_{\rm YM}$,
and $\Sigma_{IJK}$ is a $d_{IJK}$-dimensional subcycle over which the three zero mode wavefunctions $\phi_I^{(0)}, \phi_J^{(0)}$ and $\phi_K^{(0)}$
 can have a nonzero overlap.
Here $\phi_I^{(0)}$ is 
 is normalized as 
\bea
\int_{\Sigma_I} |\phi_{I}^{(0)}|^2 \, d{\cal V}(\Sigma_I) =1,\eea
Note that \bea
\Sigma_{\rm YM}\supseteq \Sigma_I \supseteq \Sigma_{IJK},\eea and therefore \bea
d_{\rm YM}\geq d_I\geq d_{IJK}.\eea

It has been argued in \cite{hep-th/0609180} that in the limit where ${\cal V}(\Sigma_{\rm YM})$ is large enough and therefore the flux densities in $\Sigma_{\rm YM}$ can be ignored,
 the normalized zero mode wavefunction
$\phi_I^{(0)}$ is simply rescaled as
\bea
  \phi_I^{(0)}\rightarrow \lambda^{-d_I/2d_{\rm YM}}\phi_I^{(0)}\eea
under the rescaling 
\bea
\label{vol_re}
{\cal V}(\Sigma_{\rm YM})\rightarrow
\lambda {\cal V}(\Sigma_{\rm YM}). \eea
Then
the associated Yukawa couplings are rescaled as  
\bea
\label{yukawa_scaling}
y_{IJK} \rightarrow \lambda^{(2d_{IJK}-d_I-d_J-d_K)/2d_{\rm YM}}y_{IJK}.\eea
On the other hand,
  the physical Yukawa couplings in 4-dimensional supergravity  are given by
\bea
\label{yukawa_sugra}
y_{IJK} =\frac{\lambda_{IJK}}{\sqrt{Y_IY_JY_K}},\eea
where $\lambda_{IJK}$ are the holomorphic Yukawa couplings in the superpotential:
\bea
W_{\rm Yukawa}=\frac{1}{6}\lambda_{IJK}\Phi^I\Phi^J\Phi^K\eea
and
\bea Y_I=e^{-K_0/3} Z_I.\eea
Due to the holomorphicity and the $U(1)_{\rm PQ}$ symmetry, $\lambda_{IJK}$ are 
independent of ${\cal V}(\Sigma_{\rm YM})$.
Since ${\rm Re}({\cal F}_A)=1/g_A^2 \propto {\cal V}(\Sigma_{\rm YM})$, 
 (\ref{yukawa_scaling}) and (\ref{yukawa_sugra})
 suggest 
\bea\label{scaling_relation}
Y_I=e^{-K_0/3}Z_I\propto \Big({\rm Re}({\cal F}_A)\Big)^{\omega_I},\eea
where the scaling weights  $\omega_I$ are rational numbers constrained as
\bea
\label{weight_relation}
\quad \omega_I+\omega_J+\omega_K=\frac{1}{d_{\rm YM}}(d_I+d_J+d_K-2d_{IJK})\eea
for the combinations having a nonzero Yukawa coupling. 
This implies
\bea
\frac{\partial \ln Y_I}{\partial T}\simeq {\omega_I}\frac{\partial \ln {\rm Re}({\cal F}_A)}{\partial T}=\omega_I\frac{c_Ag_A^2}{16\pi^2}
\eea
which confirms  the estimates 
(\ref{estimate0}) and  (\ref{estimate}).
Although the above scaling argument is valid only 
at leading order  in $1/{\cal V}(\Sigma_{\rm YM})$, 
it is likely to give a correct order of magnitude estimate even when higher order corrections  become non-negligible.


An interesting string model which can incorporate a QCD axion
is the Large Volume Scenario (LVS) \cite{hep-th/0502058} of Type IIB string 
compactification on a Calabi-Yau (CY) space which 
involves a 4-cycle wrapped by $D7$ branes  supporting the SM gage fields.  It has been argued that in this scenario  
all moduli including the saxion partner of the QCD axion 
can be successfully stabilized  with
$f_a$ exponentially lower than the Planck scale \cite{hep-th/0602233,1206.0819}.
At leading order approximation to ignore the effects of magnetic fluxes on $D7$ branes,
the Kahler potential and the holomorphic gauge kinetic function of the model are given by \cite{hep-th/0609180, hep-th/0602233}
\bea
\label{lvs}
K=-3\ln(T_b+T_b^*) + \frac{(T+T^*)^{3/2}}{(T_b+T_b^*)^{3/2}} +\frac{(T+T^*)^{\omega_I}}{(T_b+T_b^*)}\Phi_I^*\Phi_I,\quad
{\cal F}_A=\frac{1}{8\pi^2}T,\eea
where 
\bea
T_b=\tau_b+i\theta_b, \quad T=\tau +i\frac{a}{f_a}  \eea
and the scaling weights $\omega_I$ are rational numbers between 0 and 1 as suggested by (\ref{weight_relation}).
Note that the factor $1/8\pi^2$ in ${\cal F}_A$ is due to the normalization convention of  ${\rm Im}(T)=a(x)/f_a\cong {\rm Im}(T)+2\pi$, where $a(x)$ is the QCD axion in the model. 
Here the Kahler modulus $\tau_b$ corresponds to the volume of an exponentially
big  4-cycle in CY space, with which the 4-dimensional Planck scale $M_P$ can be exponentially bigger than the 
string scale $M_{\rm st}$ as
\bea
\tau_b \sim {\cal V}_{\rm CY}^{2/3}\sim \left(\frac{M_P}{M_{\rm st}}\right)^{4/3} \gg  
\frac{8\pi^2}{g_{\rm GUT}^2},\eea
where  ${\cal V}_{\rm CY}$ is the CY volume,
 while $\tau$ describes the volume of a moderately large 4-cycle wrapped by
$D7$ branes supporting the SM gauge fields, which determines the SM gauge couplings at the string scale as 
\bea
\tau =\frac{8\pi^2}{g_{\rm GUT}^2}.\eea
For simplicity, in (\ref{lvs}) we ignored the part depending 
on other moduli while keeping only the part relevant for the couplings and the decay constant of the QCD axion $a(x)$. 
We then find 
\bea
f_a = {\frac{\sqrt{3}}{2\sqrt{2}}}\frac{M_P}{\tau^{1/4} \tau_b^{3/4} }\eea
 together with
the following  couplings  to gauge and matter fields: \bea
 c_A=1, \quad
  c_{\phi_I}(f_a)=\omega_I \frac{g_{\rm GUT}^2}{16\pi^2}+{\cal O}\Big(\frac{1}{\tau_b}\Big) ,\quad c_{\psi_I} =\omega_I \frac{g_{\rm GUT}^2}{16\pi^2}+{\cal O}\Big(\frac{1}{\tau_b}\Big). \eea
{This model of string-theoretic axion is particularly interesting as it can give rise to an axion decay constant 
anywhere between  $f_a\sim 10^{16}$ GeV (for $\tau_b\sim \tau$) and the observational lower bound $f_a\sim 10^{9}$ GeV (for an exponentially large $\tau_b\sim 10^{11}\gg \tau$).
Therefore, in section \ref{sec:PP}, we use this model as a benchmark model for string-theoretic axion and compare its low energy couplings $g_{aX}$ ($X=\gamma, p, n, e$) with those of KSVZ-like axions. More    
specifically, we will consider a model 
with chiral matters on the intersections of $D7$ branes, for which
 $d_{\rm YM}=4$, $d_I=2$ and $d_{IJK}=0$, and therefore \cite{hep-th/0609180}  
\bea \label{sw1/2}
\omega_I=\frac{1}{2}.\eea}

In order to see that the order of magnitude estimate (\ref{estimate}) is valid even when the flux densities  in $\Sigma_{\rm YM}$ are not negligible, let us consider another example of string-theoretic axions, the model-independent
axion and the model-dependent Kahler axion in heterotic string or $M$-theory 
 in which the underlying YM and gravitational fluxes play a key role
for the model to accommodate a QCD axion \cite{Witten:1984dg, Choi:1985bz,hep-th/9605136, hep-th/9706171}.
For simplicity, we consider a compactification on CY space with a single Kahler axion, 
which involves
\bea
S=\tau_1+i\frac{a_1}{f_1}, \quad  T=\tau_2+i\frac{a_2}{f_2},
\eea
where $a_1\cong a_1+2\pi f_1$ and $a_2\cong a_2+2\pi f_2$ are
 the model-independent axion  and the Kahler axion, respectively, that 
originate from the NS 2-form field $B$ as
\bea
\partial_\mu a_1\propto  \epsilon_{\mu\nu\rho\sigma}\partial^\nu B^{\rho\sigma},
\quad a_2 \propto  \int_{\Sigma} B_{i\bar j}dz^i \wedge dz^{*j},\eea
where $\Sigma$ is the 2-cycle dual to the Kahler form.
In the heterotic string regime, $\tau_1$ correspond to the Euclidean action of the NS5 brane instanton wrapping the entire CY space \cite{Polchinski:1998rr} , and $\tau_2$ is that  of the world-sheet instanton wrapping $\Sigma$ \cite{Dine:1986zy}:
\bea
\tau_1\propto \frac{{\cal V}_{\rm CY}}{g_{\rm st}^2}, \quad \tau_2\propto {\cal V}({\Sigma}), \eea
where $g_{\rm st}$ is the string coupling. The moduli $\tau_1$ and $\tau_2$ can be extrapolated to the heterotic $M$-theory regime \cite{hep-th/9603142} where they correspond to the Euclidean actions of the $M5$ brane instanton wrapping the CY space and the $M2$ brane instanton wrapping 
$\Sigma$ stretched along the 11-th dimension.

The moduli Kahler potential, matter Kahler metric, and holomorphic gauge kinetic functions of the above model are given by \cite{hep-th/9710208}
\bea
K_0&=&-\ln(S+S^*)-3\ln(T+T^*), \nonumber \\
Z_I &=&\frac{1}{T+T^*}-\frac{\epsilon}{3}\frac{1}{S+S^*},\nonumber \\ 
{\cal F}_{E6} &=& \frac{1}{8\pi^2}(S+\epsilon T), \quad {\cal F}_{E8}\, =\,\frac{1}{8\pi^2}(S-\epsilon T)\eea
where {${\cal F}_{E6}$ and ${\cal F}_{E8}$ are the gauge kinetic functions of the visible $E_6$ and the hidden $E_8$, respectively, and} 
\bea {\epsilon={\frac{1}{8\pi^2}}\int J\wedge \left[{\rm tr}(F\wedge F) -\frac{1}{2}{\rm tr}(R\wedge R)\right]}\eea 
for the Kahler form $J$. 
Obviously the pieces proportional to $\epsilon$ are the corrections induced by flux density,
which can be interpreted as string loop corrections in the heterotic string regime
\cite{Choi:1985bz, Derendinger:1985cv, Ibanez:1986xy}, or as
a flux-induced warping of CY geometry along the 11-th dimension   in the heterotic $M$-theory regime
\cite{hep-th/9602070, hep-th/9710208}.
At leading order approximation which ignores these corrections, which would correspond to the weakly coupled heterotic string  theory limit with $\epsilon \tau_2\ll \tau_1$, one finds 
\bea
Y_I =e^{-K_0/3}Z_I \simeq (S+S^*)^{1/3} \propto \Big({\rm Re}({\cal F}_A)\Big)^{1/3}\eea
as suggested by the scaling relation (\ref{scaling_relation}) with $\omega_I=1/3$ which is
implied by (\ref{weight_relation}) for gauge and matter fields propagating over the full CY space:
$d_{\rm YM}=d_I=d_{IJK}=6$.
However, due to the hidden $E_8$ gauge group confining at high scales as well as the world sheet instanton effects, a viable QCD axion can be obtained {\it only} in the heterotic $M$-theory limit with $\epsilon \tau_2\sim \tau_1={\cal O}(8\pi^2/g_{\rm GUT}^2)$ \cite{hep-th/9605136, hep-th/9706171}.
In such case, the QCD axion $a(x)$ corresponds to a combination of $a_1$ and $a_2$ which avoids the coupling to the hidden $E_8$ gauge fields.
To find the decay constant and couplings of this QCD axion, one may parameterize the model-independent
axion and the Kahler axion in terms of the QCD axion as
\bea
\frac{a_1}{f_1} \propto {\epsilon \frac{a}{f_a}}, \quad \frac{a_2}{f_2} \propto \frac{a}{f_a},\eea
where the heavy axion component is frozen at its vacuum value by the potential induced by the hidden $E_8$ gauge fields.
For the convenience of  discussion, here we define $f_a$ as a scale for which 
the couplings of $a(x)$ to  the $E_6$ gauge fields are given by
$c_A=1$, {\it not} as a scale to define the periodicity of $a(x)$.
It is then straightforward to find 
\bea
f_a =  \sqrt{2} M_P \frac{g_{\rm GUT}^2 }{8\pi^2} \frac{g_{\rm hid}^2 \sqrt{(g_{\rm hid}^2 + g_{\rm GUT}^2)^2-g_{\rm hid}^2 g_{\rm GUT}^2}}{|g_{\rm hid}^4-g_{\rm GUT}^4|}\eea
with the couplings
\bea
 c_A=1, \quad
  c_{\phi_I}(f_a)=\omega_{\phi} \frac{g_{\rm GUT}^2}{16\pi^2} ,\quad c_{\psi_I} =\omega_{\psi} \frac{g_{\rm GUT}^2}{16\pi^2}, \eea
  where
  \bea
  \omega_\phi&=& -\frac{g_{\rm hid}^2\left[(g_{\rm hid}^2 + g_{\rm GUT}^2)^2 + 2 g_{\rm hid}^2 g_{\rm GUT}^2\right]}{\left( g_{\rm hid}^2 + 2 g_{\rm GUT}^2\right)\left(g_{\rm hid}^4 - g_{\rm GUT}^4\right)},\nonumber \\
  \omega_\psi&=& \frac{g_{\rm hid}^2\left[(g_{\rm hid}^2 + g_{\rm GUT}^2)^2 -  g_{\rm hid}^2 g_{\rm GUT}^2\right]}{\left( g_{\rm hid}^2 + 2 g_{\rm GUT}^2\right)\left(g_{\rm hid}^4 - g_{\rm GUT}^4\right)},
 \eea
  and { $g_{\rm GUT}$ and $g_{\rm hid}$ are the gauge couplings around the string scale for the visible $E_6$ and the hidden $E_8$, respectively, which are determined by}
  \bea
  {\rm Re}({\cal F}_{E_6}) =\frac{1}{g_{\rm GUT}^2}, \quad {\rm Re}({\cal F}_{E_8})=\frac{1}{g_{\rm hid}^2}.\eea
This result shows that the order of magnitude estimate (\ref{estimate}) is valid even when the effects of flux densities  are properly taken into account. 

\section{Running of the axion couplings \label{sec:RG}}

The low energy axion couplings can be substantially different from the UV boundary values discussed in the previous section 
because of the subsequent  renormalization group evolution.
In this section we will present the RG equations of axion couplings to the SM particles at leading order in dimensionless couplings (i.e. Yukawa and gauge couplings) in a generic way that can be applied to the SM, the minimal supersymmetric standard model (MSSM), and the two Higgs doublet models (2HDMs). 
We will then provide a semi-analytic formula for the solution of the RG equations as well as the numerical results.

\subsection{RG equations \label{sec:RGeq}}

In the Georgi-Kaplan-Randall (GKR) field basis \cite{Georgi:1986df},
the axion couplings to the SM fields and an additional Higgs doublet field in supersymmetric models or 2HDMs can be generally written
\dis{
\label{coup_2hdm}
 {\cal L}_a =\frac{\partial_\mu a}{\vpq}\left[\,\sum_{\psi} (\C_\psi)_{ij} \psi_i^\dagger \bar{\sigma}^\mu \psi_j
 +\sum_{\alpha=1,2} c_{H_\alpha}H_\alpha^\dagger\overset{\leftrightarrow}{i D^\mu} H_\alpha\,\right] + \frac{a}{\vpq} \sum_{A} c_A \frac{g_A^2}{32 \pi^2} F^{A \mu \nu} \widetilde{F}^A_{\mu \nu},
}
where $\psi_i = \{ Q_i, u_i^c, d_i^c, L_i, e_i^c\} \, (i=1,2,3)$ are the left-handed quarks and leptons in the SM and $H_\alpha$ ($\alpha=1,2$) denote the two Higgs doublets in the model.
\begin{figure}  
\centering
\includegraphics[scale=0.065]{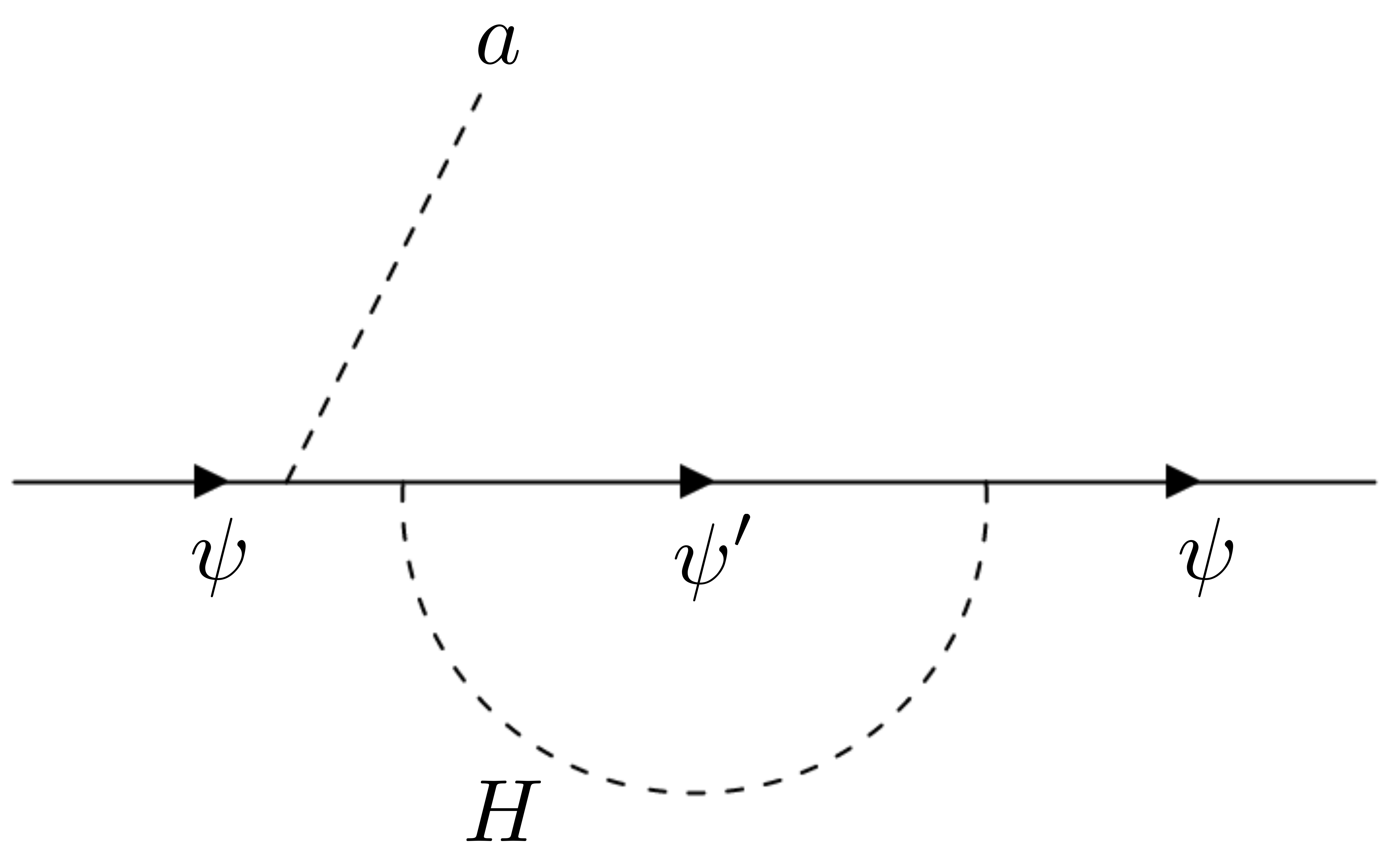}
\includegraphics[scale=0.065]{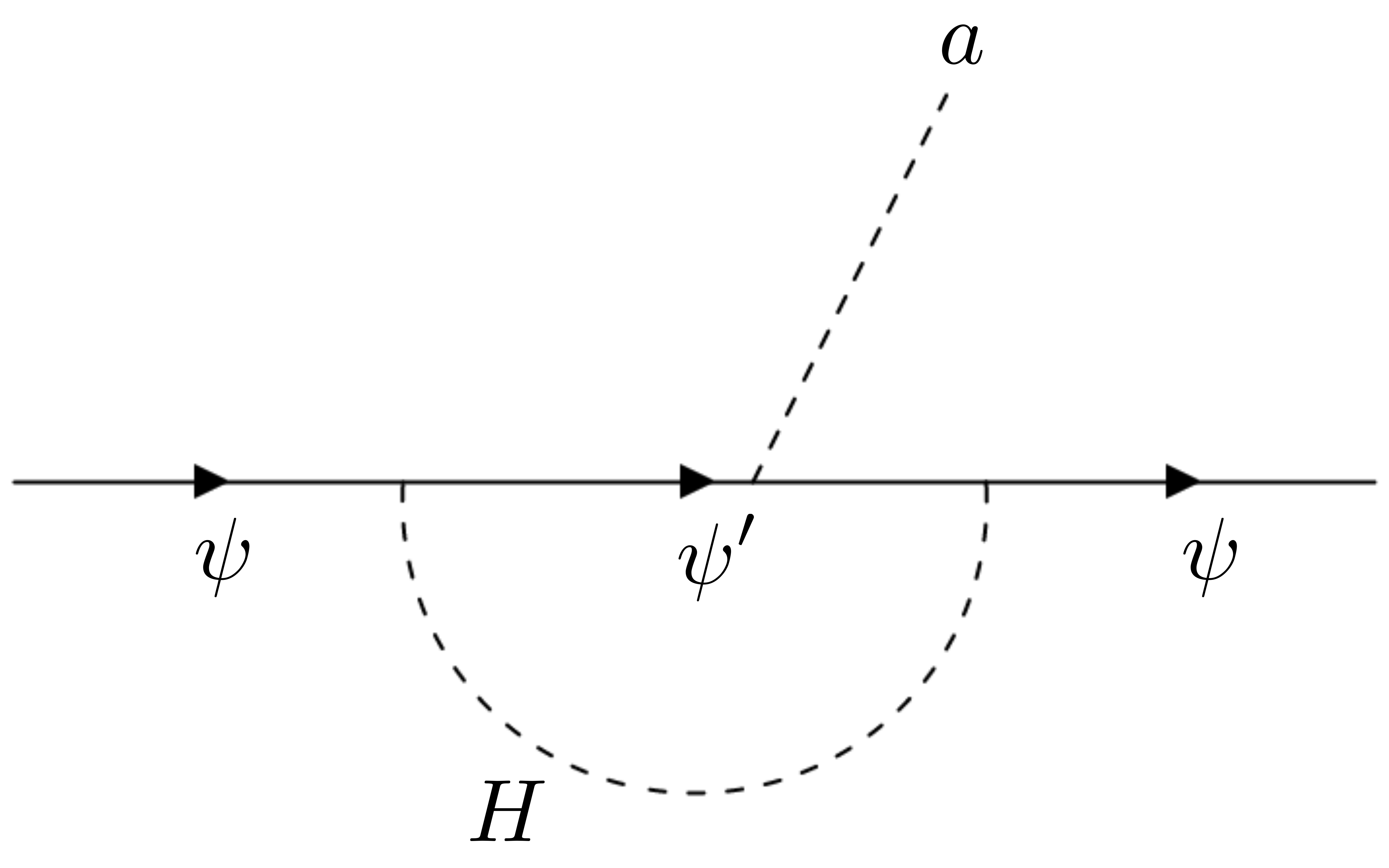}
\includegraphics[scale=0.065]{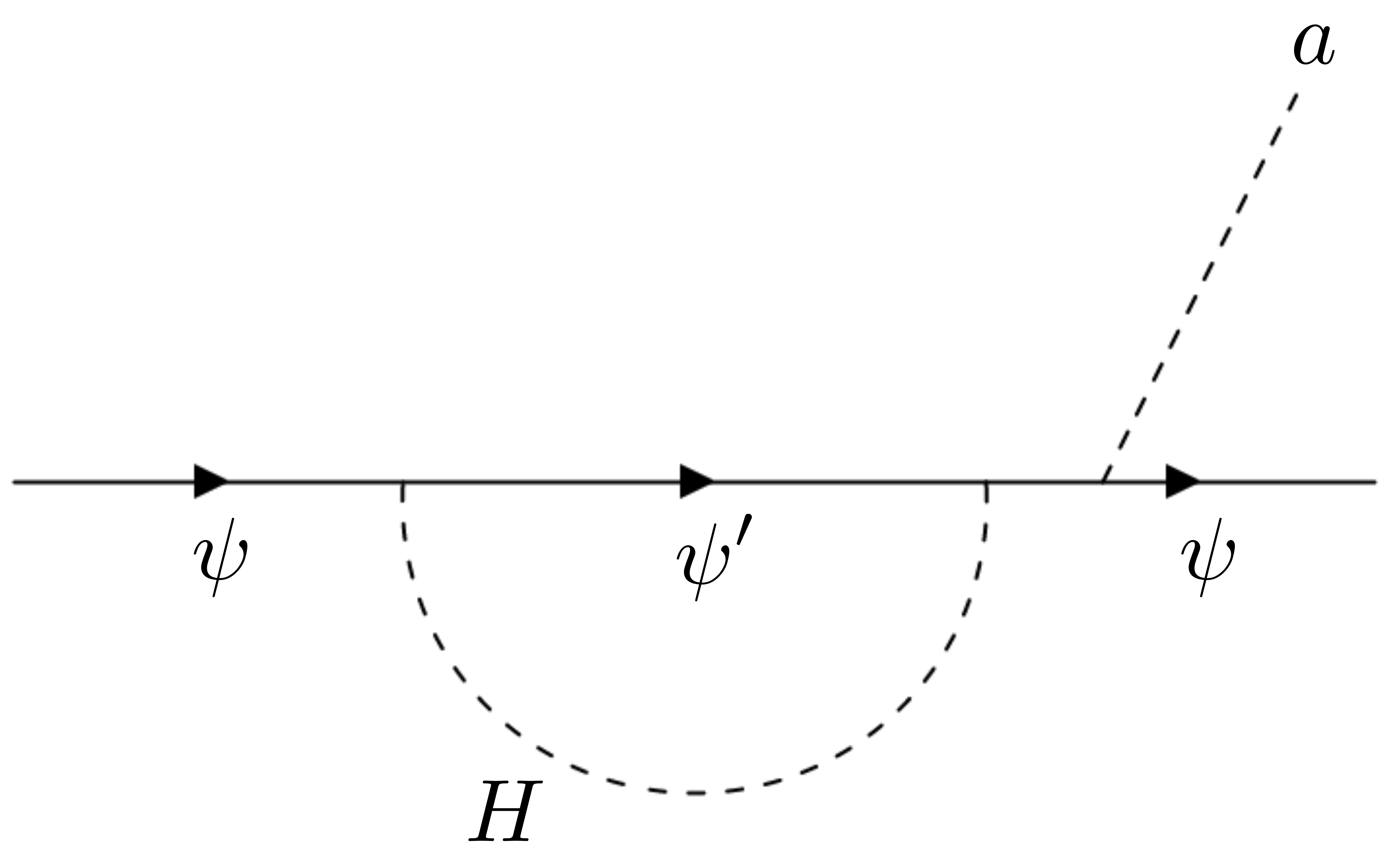} 
\includegraphics[scale=0.065]{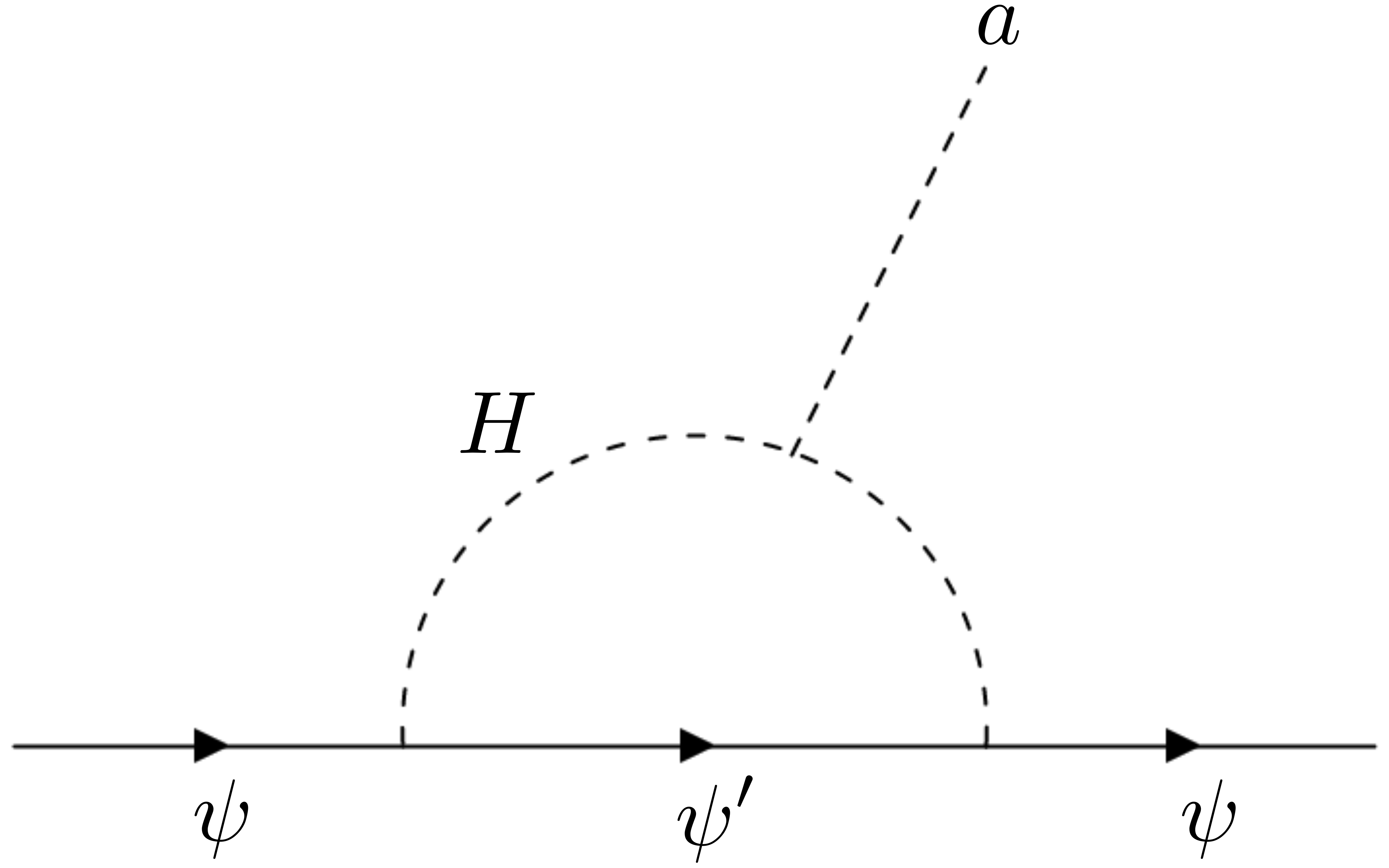}
\caption{Diagrams for the Yukawa-induced  renormalization group running of $\C_\psi$.}
\label{fig:yukawa}
\end{figure} 
The axion couplings to $\psi_i$ and $H_\alpha$  get renormalized by the Yukawa interactions involving 
$H_\alpha$ and $\psi_i$, which is determined at leading order by the diagrams in Fig. \ref{fig:yukawa}.
For general 2HDMs, the Yukawa couplings can be written as
\dis{ \label{yukawa_2hdm}
{\cal L}_{\rm Yukawa}=\sum_{f,F,\alpha, i, j} (\y_{fF\alpha})_{ij} f_i F_ j H_\alpha +\textrm{h.c.},
}
where $f_i = \{u^c_i, d^c_i, e^c_i \}$ are $SU(2)_L$-singlet fermions, and $F_i = \{ Q_i, L_i \}$ are $SU(2)_L$-doublet fermions. 
The  Yukawa-induced RG running of axion couplings at leading order are previously given in \cite{1708.00021, 2002.04623, 2011.08205, 2012.09017, 2012.12272} for the SM or its supersymmetric extensions.
Here we provide a general expression which can be applied to general 2HDMs as well:
\bea 
 \label{RG-yukawa}
\left.\frac{d\C_F}{d\ln \mu}\right|_{\rm Yukawa} & = & \frac{\xi_y}{16 \pi^2}\sum_{f, \alpha} \left( \frac12 \{\C_F, \y_{fF\alpha}^{\dagger}  \y_{fF\alpha}  \}
 + \y_{fF\alpha}^\dagger \C_{f}^T   \y_{fF\alpha} + c_{H_\alpha}\y_{fF\alpha}^\dagger \y_{fF\alpha}  \right), \nonumber \\
\left.\frac{d\C_{f}^T}{d\ln \mu}\right|_{\rm Yukawa} & = & \frac{\xi_y}{16\pi^2}\sum_{F, \alpha} \left( \frac12 \{\C_{f}^T , \y_{fF\alpha} \y_{fF\alpha}^\dagger\} + \y_{fF\alpha} \C_F \y_{fF\alpha}^\dagger + c_{H_\alpha}\y_{fF\alpha} \y_{fF\alpha}^\dagger \right),  \nonumber \\
\left.\frac{d c_{H_\alpha}}{d\ln \mu}\right|_{\rm Yukawa} & = & \frac{1}{8\pi^2}\sum_{f, F} \left( c_{H_\alpha} {\rm tr}(\y_{fF\alpha}^\dagger \y_{fF\alpha}) + {\rm tr}(\y_{fF\alpha} \C_F \y_{fF\alpha}^\dagger) + {\rm tr}(\y_{fF\alpha}^\dagger \C_{f}^T \y_{fF\alpha})  \right),  \nonumber \\
\eea
where 
\dis{
\xi_y = 
\begin{cases} 1  &\textrm{for non-SUSY models} \\
2 & \textrm{for SUSY models}
\end{cases}
}

\begin{figure}  
\centering
\includegraphics[scale=0.3]{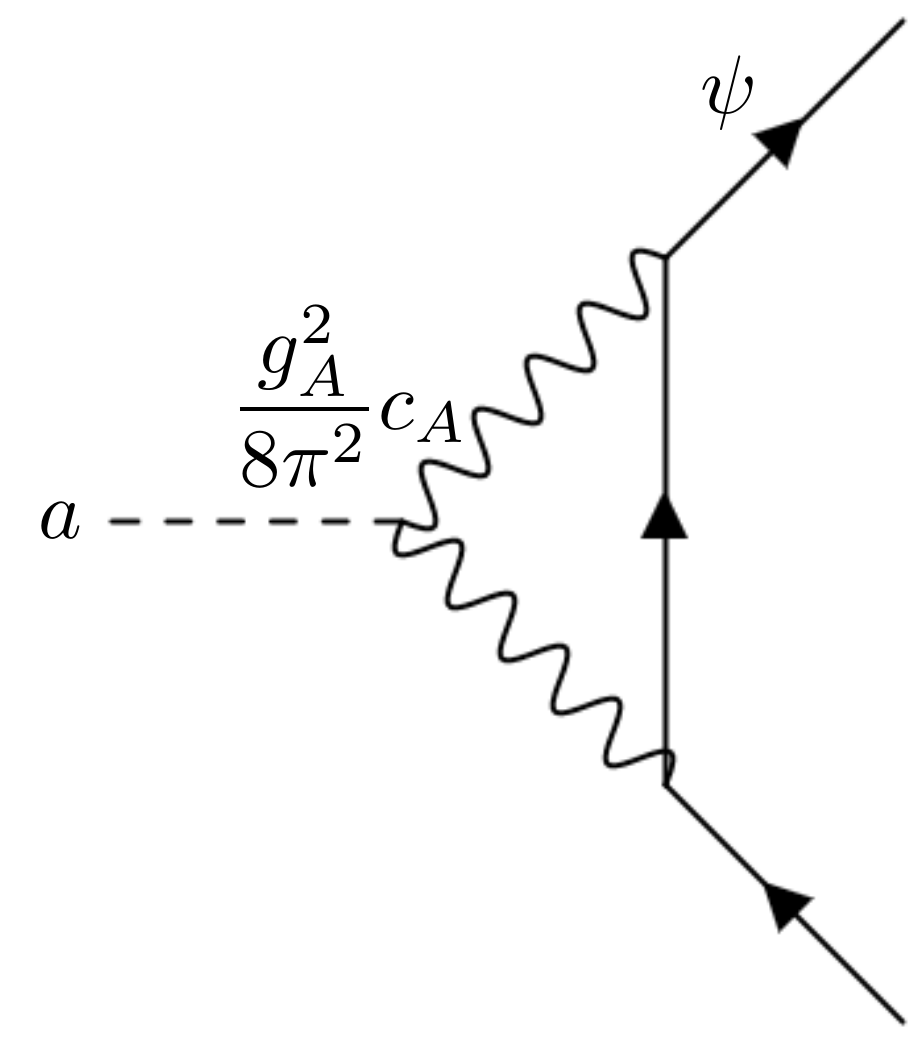} \hspace{0.7cm}
\includegraphics[scale=0.3]{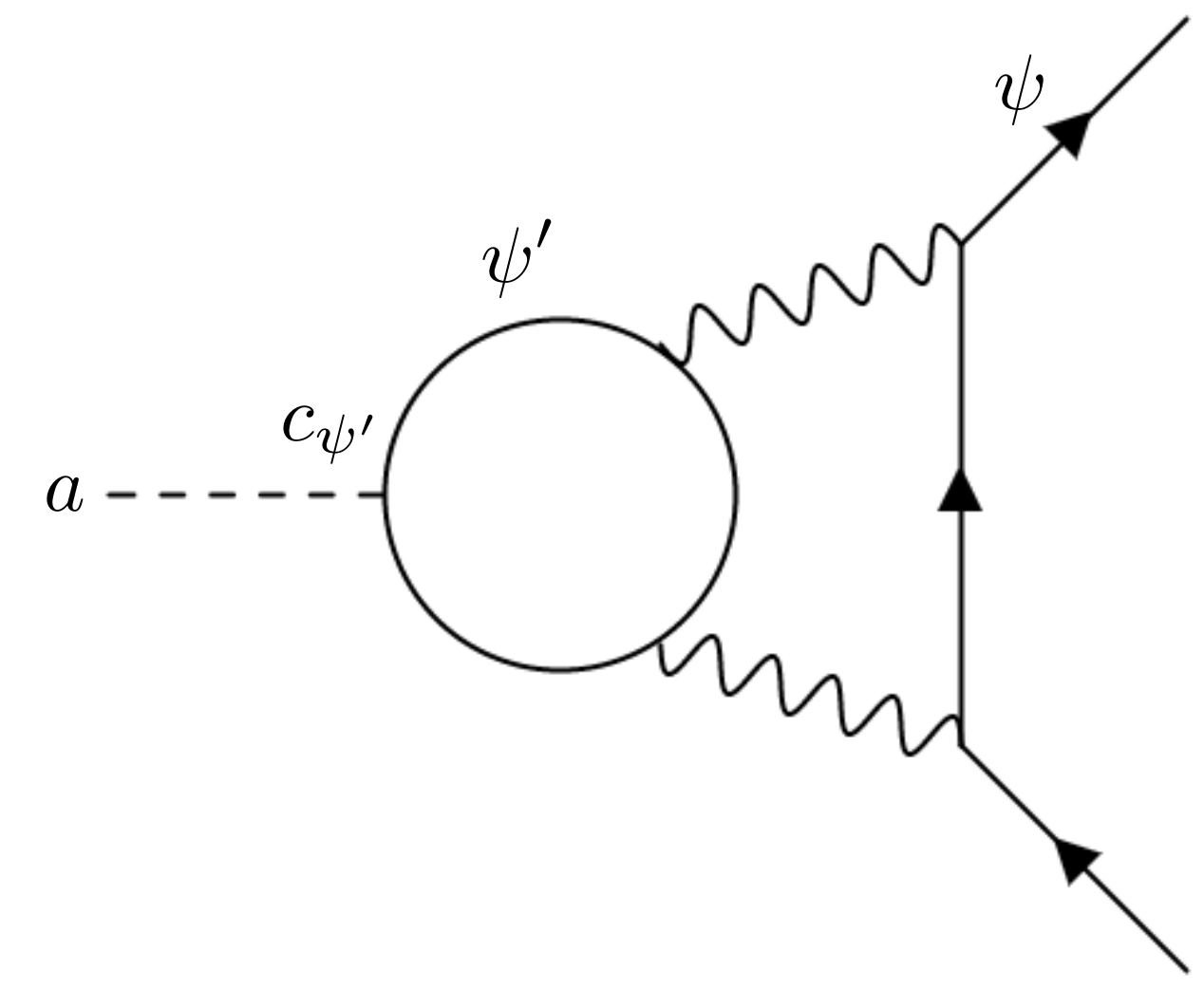}
\caption{Diagrams for the gauge-induced  renormalization group 
running of $\C_\psi$.}
\label{fig:gauge}
\end{figure}
Gauge interactions also give rise to a running of axion couplings which is determined at leading order  by the diagrams in Fig. \ref{fig:gauge}.
A general formula for such gauge-induced RG running of axion couplings is given in \cite{2012.05029, 2012.09017, 2012.12272} for the SM. 
Here we generalize the formula for SUSY models using a connection between the beta function of an axion coupling and the anomalous dimension of a chiral superfield  as described in \cite{1708.00021, ArkaniHamed:1998kj}:
\bea
\label{RG-gauge}
\left.\frac{d \C_\psi}{d \ln \mu}\right|_{\rm gauge} &=&  -\xi_g\sum_{A} \frac{3}{2}\left(\frac{g_A^2}{8\pi^2}\right)^2\mathbb{C}_{A}(\psi)\Big( c_A - 2\sum_{\psi'}
{\rm tr} (\C_{\psi^\prime})\mathbb{T}_{A}(\psi')  \Big) \mathbb{1}, \nonumber \\
\left.\frac{d c_{H_\alpha}}{d \ln \mu}\right|_{\rm gauge} &=&  -\xi_H\sum_{A} \frac{3}{2}\left(\frac{g_A^2}{8\pi^2}\right)^2\mathbb{C}_{A}(H_\alpha)\Big( c_A - 2\sum_{\psi'}
{\rm tr} (\C_{\psi^\prime})\mathbb{T}_{A}(\psi')  \Big), 
\eea
where
\dis{
\xi_g = 
\begin{cases} 1 &\textrm{for non-SUSY models} \\
2/3 & \textrm{for SUSY models}
\end{cases}, \quad
\xi_H = 
\begin{cases} 0 &\textrm{for non-SUSY models} \\
2/3 & \textrm{for SUSY models}
\end{cases}
}
The difference between non-SUSY models and SUSY modes is due to the presence of the axion-gaugino couplings in SUSY models. Here $\mathbb{C}_{A}(\Phi)$ and $\mathbb{T}_{A}(\Phi)$ are the quadratic Casimir and Dynkin index  of the field $\Phi$ charged under the gauge group $A$.

In fact, Eq. (\ref{RG-yukawa}) and Eq. (\ref{RG-gauge}) can be determined up to overall coefficients by 
the covariance of the RG equations under two spurious symmetries of the lagrangian. 
One of the symmetries is associated with the $SU(3)$ flavor rotations of $\psi_i$: 
\dis{
\psi_i \,\rightarrow\, (U_\psi)_{ij} \psi_j, \quad \C_\psi \,\rightarrow\, U_\psi \C_\psi U_\psi^\dagger, \quad
\y_{f F\alpha} \,\rightarrow\, U_f^* \y_{fF\alpha} U_F^\dagger.
}
The other involves the following axion-dependent field redefinition which is allowed within the GKR field basis: 
\dis{ \label{repar}
\psi_i \rightarrow \psi_i e^{i x_{\psi_i} a/f_a}, \quad H_\alpha \rightarrow H_\alpha e^{i x_{H_\alpha} a/f_a} 
}
for $x_{\psi_i}$ and $x_{H_\alpha}$ satisfying
\dis{
(\y_{fF\alpha})_{ij} \left(x_{f_i} + x_{F_j} + x_{H_\alpha}\right)=0, \nonumber
}
with redefinition of parameters
\dis{
(\C_\psi)_{ij} \, \rightarrow\, (\C_\psi)_{ij} + x_{\psi_i} \delta_{ij}, \quad c_{H_\alpha} \, \rightarrow\, c_{H_\alpha} + x_{H_\alpha}, \quad c_{A} \, \rightarrow\, c_A + 2 \sum_{\psi_i} x_{\psi_i} \mathbb{T}_A(\psi_i). \\
}

From the RG equations (\ref{RG-yukawa}) and (\ref{RG-gauge}), 
we find that the condition for a non-trivial RG running of the couplings of field-theoretic axions can be expressed in terms of the linearly realized $U(1)_{\rm PQ}$ symmetry. 
For field-theoretic axions, the axion couplings to matter fields in the GKR field basis at $\mu\sim f_a$ are identified as the PQ charges of those fields in the field basis where $U(1)_{\rm PQ}$ is linearly realized as Eq. (\ref{linear_pq}), while the axion couplings  to gauge fields  in the GKR field basis is identified as the $U(1)_{\rm PQ}$ anomaly for the gauge group. Then, from Eq. (\ref{RG-yukawa}), one can see that the RG running of axion couplings induced by a Yukawa coupling occurs when the sum of the PQ charges of the fields involved in the corresponding Yukawa coupling is non-vanishing. On the other hand, Eq. (\ref{RG-gauge}) tells us that the RG running of axion couplings induced by  gauge interaction starts from the scale where the heaviest PQ- and gauge-charged fermion is integrated out.

In the following analysis we will focus on the RG evolution of axion couplings in the MSSM and the SM. We write the Yukawa couplings in the MSSM as
\dis{
{\cal L}_{\rm Yukawa}=(\Y_u)_{ij} {u^c_i}  Q_{j} H_u + (\Y_d)_{ij}  {d^c_i} Q_{j} H_d + (\Y_e)_{ij} {e^c_i}  L_j H_d + \textrm{h.c.}
}
From the general formula Eq. (\ref{RG-yukawa}) and Eq. (\ref{RG-gauge}), the full axion RG equations at leading order in the Yukawa and gauge couplings in the MSSM come out as
\bea \label{RG-MSSM}
\frac{d\C_Q}{d\ln \mu} & = & \frac{\xi_y}{16 \pi^2}\left( \frac12 \{\C_Q, \Y_u^\dagger  \Y_u + \Y_d^\dagger \Y_d \}
 + \Y_u^\dagger \C_{u^c}^T   \Y_u + \Y_d^\dagger \C_{d^c}^T \Y_d + c_{H_u}\Y_u^\dagger \Y_u + c_{H_d}\Y_d^\dagger \Y_d  \right)  \nonumber \\
 &&-\xi_g\left(\frac{\alpha_s^2}{2\pi^2} \tilde{c}_G  +\frac{9\alpha_2^2}{32\pi^2} \tilde{c}_W +\frac{\alpha_1^2}{96\pi^2} \tilde{c}_B\right) \mathbb{1}\,, \nonumber \\
\frac{d\C_{u^c}^T}{d\ln \mu} & = & \frac{\xi_y}{8\pi^2}\left( \frac12 \{\C_{u^c}^T , \Y_{u} \Y_{u}^\dagger\} + \Y_{u} \C_Q \Y_{u}^\dagger + c_{H_u}\Y_u \Y_u^\dagger \right) -\xi_g\left(\frac{\alpha_s^2}{2\pi^2} \tilde{c}_G  +\frac{\alpha_1^2}{6\pi^2} \tilde{c}_B\right) \mathbb{1}\,, \nonumber \\
\frac{d\C_{d^c}^T}{d\ln \mu} & = & \frac{\xi_y}{8\pi^2}\left( \frac12 \{\C_{d^c}^T, \Y_{d} \Y_{d}^\dagger\} + \Y_{d} \C_Q \Y_{d}^\dagger  + c_{H_d}\Y_d \Y_d^\dagger   \right)  -\xi_g\left(\frac{\alpha_s^2}{2\pi^2} \tilde{c}_G  +\frac{\alpha_1^2}{24\pi^2} \tilde{c}_B\right) \mathbb{1}\,, \nonumber  \\
\frac{d\C_L}{d\ln \mu} & = & \frac{\xi_y}{16\pi^2}\left(  \frac12 \{ \C_L, \Y_e^\dagger \Y_e\} +\Y_e^\dagger \C_{e^c}^T \Y_e  +c_{H_d}\Y_e^\dagger \Y_e \right) -\xi_g\left(\frac{9\alpha_2^2}{32\pi^2} \tilde{c}_W +\frac{3\alpha_1^2}{32\pi^2} \tilde{c}_B\right) \mathbb{1}\,, \nonumber  \\
\frac{d\C_{e^c}^T}{d\ln \mu} & = & \frac{\xi_y}{8\pi^2}\left(\frac12 \{\C_{e^c}^T, \Y_{e} \Y_{e}^\dagger\} + \Y_{e} \C_L \Y_{e}^\dagger  + c_{H_d}\Y_e \Y_e^\dagger \right)  -\xi_g \frac{3\alpha_1^2}{8\pi^2} \tilde{c}_B \mathbb{1}\,, \nonumber  \\
\frac{d c_{H_u}}{d\ln \mu} & = & \frac{3}{8\pi^2}\left( c_{H_u} {\rm tr}(\Y_u^\dagger \Y_u) + {\rm tr}(\Y_u \C_Q \Y_u^\dagger) + {\rm tr}(\Y_u^\dagger \C_{u^c}^T \Y_u)  \right) -\xi_H\left(\frac{9\alpha_2^2}{32\pi^2} \tilde{c}_W +\frac{3\alpha_1^2}{32\pi^2} \tilde{c}_B\right), \nonumber  \\
\frac{d c_{H_d}}{d\ln \mu} & = & \frac{3}{8\pi^2}\left( c_{H_d} {\rm tr}(\Y_d^\dagger \Y_d) + {\rm tr}(\Y_d \C_Q \Y_d^\dagger) + {\rm tr}(\Y_d^\dagger \C_{d^c}^T \Y_d) \right) \nonumber \\ 
&&+\frac{1}{8\pi^2}\left( c_{H_d} {\rm tr}(\Y_e^\dagger \Y_e) + {\rm tr}(\Y_e \C_L \Y_e^\dagger) + {\rm tr}(\Y_e^\dagger \C_{e^c}^T \Y_e)  \right) -\xi_H\left(\frac{9\alpha_2^2}{32\pi^2} \tilde{c}_W +\frac{3\alpha_1^2}{32\pi^2} \tilde{c}_B\right), \nonumber \\
\eea
with
\dis{ \label{c_gauge}
\tilde{c}_G &= c_G -{\tr\left(2 \C_Q + \C_{u^c} + \C_{d^c}\right)}\,, \\
\tilde{c}_W &= c_W -{\tr\left(3 \C_Q + \C_L\right)} - \frac{3}{2} \xi_H (c_{\widetilde{H}_u}+c_{\widetilde{H}_d}) \,, \\
\tilde{c}_B &= c_B - {\tr\left(\frac{1}{3} (\C_Q + 8 \,\C_{u^c} + 2\,\C_{d^c}) + \C_L  +2 \C_{e^c} \right)} -\frac32\xi_H( c_{\widetilde{H}_u}+c_{\widetilde{H}_d})\,.
}
The parameters $\tilde{c}_{G, W, B}$ in Eq. (\ref{c_gauge}) are invariant under the field redefinition (\ref{repar}).  
Here we note that $c_{\widetilde{H}_u}(=c_{H_u})$ and $c_{\widetilde{H}_d}(=c_{H_d})$ contributions in $\tilde{c}_W$ and $\tilde{c}_B$ are from the Higgsino loops.

In this work we are concerned with only the flavor diagonal components of the axion couplings.
{Therefore in the following we assume that $(\C_\psi)_{ij}$ are flavor diagonal in the mass eigenstate basis:\bea
(\C_\psi)_{ij}=c_{\psi_i}\delta_{ij}, 
\eea
and approximate the Yukawa couplings in the RG equations by keeping only the third generation components:
\dis{
(\Y_u)_{ij} \approx \tilde{y}_t \delta_{i3} \delta_{j3}\,, \quad (\Y_d)_{ij} \approx \tilde{y}_b \delta_{i3} \delta_{j3}\,, \quad (\Y_e)_{ij} \approx \tilde{y}_\tau \delta_{i3} \delta_{j3}\,.
}
In this approximation, the RG equations in Eq. (\ref{RG-MSSM}) can be rewritten as}
\bea
\frac{d n_{u_i}}{d\ln \mu} & \simeq &  \frac{1}{16 \pi^2}  \left( 3(2+\xi_y \delta_{i3}) \tilde{y}_t^2 n_t + \xi_y  \tilde{y}_b^2 n_b  \delta_{i3}\right) \nonumber \\
&&- \left(\frac{\xi_g}{\pi^2} \alpha_s^2 \tilde{c}_G + \frac{9(\xi_g + \xi_H)}{32\pi^2} \alpha_2^2 \tilde{c}_W +\frac{(17\xi_g + 9 \xi_H)}{96\pi^2} \alpha_1^2\tilde{c}_B \right), \nonumber  \\
\frac{d n_{d_i}}{d\ln \mu} & \simeq &  \frac{1}{16 \pi^2}  \left(\xi_y \tilde{y}_t^2 n_t  \delta_{i3} + 3(2+\xi_y \delta_{i3}) \tilde{y}_b^2 n_b + 2 \tilde{y}_\tau^2 n_\tau \right) \nonumber \\
&&- \left(\frac{\xi_g}{\pi^2} \alpha_s^2 \tilde{c}_G + \frac{9(\xi_g + \xi_H)}{32\pi^2} \alpha_2^2 \tilde{c}_W +\frac{(5\xi_g + 9 \xi_H)}{96\pi^2} \alpha_1^2\tilde{c}_B \right),  \nonumber \\
\frac{d n_{e_i}}{d\ln \mu} & \simeq &  \frac{1}{16 \pi^2}  \left(6  \tilde{y}_b^2 n_b  + (2+3\xi_y \delta_{i3})  \tilde{y}_\tau^2 n_\tau \right) -\left( \frac{9(\xi_g + \xi_H)}{32\pi^2} \alpha_2^2 \tilde{c}_W +\frac{3(5\xi_g +  \xi_H)}{32\pi^2} \alpha_1^2\tilde{c}_B\right),  \nonumber \\ \nonumber \\
\frac{d n_{H}}{d\ln \mu} & \simeq &  \frac{1}{8 \pi^2}  \left( 3 \tilde{y}_t^2 n_t + 3  \tilde{y}_b^2 n_b  +\tilde{y}_\tau^2 n_\tau \right) -\xi_H\left( \frac{9}{16\pi^2} \alpha_2^2 \tilde{c}_W +\frac{3}{16\pi^2} \alpha_1^2\tilde{c}_B \right). \label{RG-npar}
\eea
where $u_i = (u, c, t)$, $d_i = (d, s, b)$,  $e_i = (e, \mu, \tau)$, and
{
\dis{ \label{npar}
n_{u_i} &\equiv c_{Q_i} + c_{u^c_i} + c_{H_u}, \quad
 n_{d_i} \,\equiv c_{Q_i} + c_{d^c_i} + c_{H_d}, \\
 n_{e_i} &\equiv c_{L_i} + c_{e^c_i} + c_{H_d}, \quad
 n_H \,\equiv c_{H_u} + c_{H_d}.
 }
 }
Note that the parameters defined in Eq. (\ref{npar}) are invariant under the field redefinition (\ref{repar}), so they may be directly related to physically observable quantities as we will see below.
In Eq. (\ref{RG-npar}), the RG coefficients $\tilde{c}_G, \tilde{c}_W$ and $\tilde{c}_B$ are scale-dependent because the axion-fermion couplings in Eq. (\ref{c_gauge}) are running, while
\dis{ \label{cGWB}
\frac{d c_{A}}{d \ln \mu} = 0 \quad (A=G, W, B).
}
From Eq. (\ref{cGWB}) and Eq. (\ref{RG-MSSM}), one can derive
\bea \label{RG-ctilde}
\frac{ d \tilde{c}_G } { d \ln \mu} &\simeq& - \frac{\xi_y}{4\pi^2} \left(\tilde{y}_t^2 n_t + \tilde{y}_b^2 n_b \right) + \xi_g\left(\frac{6\alpha_s^2}{\pi^2} \tilde{c}_G  +\frac{27\alpha_2^2}{16\pi^2} \tilde{c}_W +\frac{11\alpha_1^2}{16\pi^2} \tilde{c}_B\right),  \nonumber \\
\frac{ d \tilde{c}_W } { d \ln \mu} &\simeq& - \frac{(\xi_y+3\xi_H)}{16\pi^2} \left(3\tilde{y}_t^2 n_t + 3\tilde{y}_b^2 n_b +\tilde{y}_\tau^2 n_\tau \right)\nonumber\\  
&&+ \frac{9\xi_g}{2\pi^2} \alpha_s^2\tilde{c}_G  +\frac{(108\xi_g + 27 \xi_H^2) }{32\pi^2} \alpha_2^2\tilde{c}_W +\frac{(12\xi_g + 9 \xi_H^2)}{32\pi^2}\alpha_1^2 \tilde{c}_B\,,  \nonumber \\
\frac{ d \tilde{c}_B } { d \ln \mu} &\simeq& - \frac{(17\xi_y+27\xi_H)}{48\pi^2}\tilde{y}_t^2 n_t - \frac{(5\xi_y+27\xi_H)}{48\pi^2}\tilde{y}_b^2 n_b - \frac{(5\xi_y+3\xi_H)}{16\pi^2} \tilde{y}_\tau^2 n_\tau \nonumber \\
 && + \frac{11\xi_g}{2\pi^2} \alpha_s^2\tilde{c}_G  +\frac{(36\xi_g + 27 \xi_H^2) }{32\pi^2} \alpha_2^2 \tilde{c}_W +\frac{(380\xi_g + 27 \xi_H^2)}{96\pi^2} \alpha_1^2\tilde{c}_B\,. 
\eea
Eq. (\ref{RG-npar}) and Eq. (\ref{RG-ctilde}) constitute a complete set of RG equations given in terms of the RG parameters invariant under the field redefinition (\ref{repar}).

The above RG equations are valid when the RG scale $\mu$ is higher than mass scales of new particles beyond the standard model (BSM). 
For simplicity, let us assume that the masses of the superparticles and charged Higgs boson mass in 2HDMs are given by a common mass scale $m_{\rm BSM}$,
\dis{
m_{\rm BSM} \sim  m_{\rm SUSY} \sim m_{H^\pm} .
}
Below the BSM scale ($\mu < m_{\rm BSM}$), the BSM particles are integrated out and no longer contribute to the running of the axion couplings. 
We will assume that $m_{\rm BSM}$ is well above the weak scale so that the light Higgs doublet field is approximated as the SM Higgs field with negligible mixing effect.
Thus the RG equations are governed by the SM degrees of freedom including the SM Higgs field $H$. As the BSM Higgs doublet field is integrated out, 
the SM Yukawa couplings $y_t, y_b$ and $y_\tau$ satisfy the matching condition
\dis{ \label{yukawa_match}
y_t = \tilde{y}_t  \sin \beta, \quad y_b = \tilde{y}_b \cos \beta, \quad y_\tau = \tilde{y}_\tau \cos \beta \quad \textrm{at} ~\mu=m_{\rm BSM},
} 
where $\tan \beta \equiv \langle H_u \rangle/ \langle H_d \rangle$,
while the coupling  between the axion and the SM Higgs field  $c_H$  satisfies
\dis{ \label{Higgs_match}
c_H = c_{H_u} \sin^2 \beta - c_{H_d} \cos^2 \beta \quad  \textrm{at} ~\mu=m_{\rm BSM}.
}
{
Then the RG equations for the axion couplings below the BSM scale (but above the weak scale)  turn out to be
\bea \label{RG-SM}
\frac{d c_{Q_i}}{d\ln \mu} & \simeq &  \frac{1 }{16 \pi^2} y_t^2 \big(c_{Q_3} + c_{u^c_3} + c_H \big)  \delta_{i3} -\left(\frac{\alpha_s^2}{2\pi^2} \tilde{c}_G  +\frac{9\alpha_2^2}{32\pi^2} \tilde{c}_W +\frac{\alpha_1^2}{96\pi^2} \tilde{c}_B\right),  \nonumber \\
\frac{dc_{u^c_i}}{d\ln \mu} & \simeq & \frac{1}{8\pi^2} y_t^2  \big(c_{Q_3} + c_{u^c_3} + c_H \big) \delta_{i3}  -\left(\frac{\alpha_s^2}{2\pi^2} \tilde{c}_G  +\frac{\alpha_1^2}{6\pi^2} \tilde{c}_B\right), \nonumber \\
\frac{d c_{d^c_i}}{d\ln \mu} & \simeq &   -\left(\frac{\alpha_s^2}{2\pi^2} \tilde{c}_G  +\frac{\alpha_1^2}{24\pi^2} \tilde{c}_B\right), \nonumber \\
\frac{d c_{L_i}}{d\ln \mu} & \simeq & -\left(\frac{9\alpha_2^2}{32\pi^2} \tilde{c}_W +\frac{3\alpha_1^2}{32\pi^2} \tilde{c}_B\right), \nonumber  \\
\frac{d c_{e^c_i}}{d\ln \mu} & \simeq &   - \frac{3\alpha_1^2}{8\pi^2} \tilde{c}_B\,,  \nonumber  \\
\frac{d c_{H}}{d\ln \mu} & \simeq & \frac{3}{8\pi^2} y_t^2 \big(c_{Q_3} + c_{u^c_3} + c_H \big),\eea
where we ignore the contributions from the bottom and tau Yukawa couplings $y_b, y_\tau$, and
\dis{ \label{c_gauge_SM}
\tilde{c}_G &= c_G -\sum_i \big(2c_{Q_i}+c_{u^c_i}+c_{d^c_i}\big)\,, \\
\tilde{c}_W &= c_W -\sum_i \big( 3c_{Q_i} +c_{L_i}\big) \,, \\
\tilde{c}_B &= c_B - \sum_i \left(\frac{1}{3}\big(c_{Q_i}+8 c_{u^c_i} +2c_{d^c_i}\big)+ c_{L_i}+2c_{e^c_i}\right)\,.
}
}
Note that there are no longer Higgsino contributions in $\tilde{c}_{W, B}$ of Eq. (\ref{c_gauge_SM}) compared to Eq. (\ref{c_gauge}).
In fact the RG equations (\ref{RG-SM}) can be recast into a simpler form:
\bea \label{RG-SM-cpar}
\frac{d C_{u_i}}{d\ln \mu} & \simeq &  \frac{3}{16 \pi^2} y_t^2 C_t (2 + \delta_{i3}) -\left(\frac{\alpha_s^2}{\pi^2} \tilde{c}_G  +\frac{9\alpha_2^2}{32\pi^2} \tilde{c}_W +\frac{17\alpha_1^2}{96\pi^2} \tilde{c}_B\right),  \nonumber \\
\frac{d C_{d_i} }{d\ln \mu} & \simeq &  -\frac{1}{16 \pi^2} y_t^2 C_t (6 - \delta_{i3}) -\left(\frac{\alpha_s^2}{\pi^2} \tilde{c}_G  +\frac{9\alpha_2^2}{32\pi^2} \tilde{c}_W +\frac{5\alpha_1^2}{96\pi^2} \tilde{c}_B\right),  \nonumber \\
\frac{d C_{e_i} }{d\ln \mu} & \simeq &    -\frac{3}{8 \pi^2} y_t^2 C_t  -\left(\frac{9\alpha_2^2}{32\pi^2} \tilde{c}_W +\frac{15\alpha_1^2}{32\pi^2} \tilde{c}_B\right).
\eea
where
{
\dis{ \label{cpar}
C_{u_i} &\equiv c_{Q_i} + c_{u^c_i} + c_{H} = n_{u_i} - n_H \cos^2 \beta, \\
 C_{d_i} &\equiv c_{Q_i} + c_{d^c_i} - c_{H}=n_{d_i} - n_H \sin^2 \beta , \\
 C_{e_i} &\equiv c_{L_i} + c_{e^c_i} - c_{H}= n_{e_i} - n_H \sin^2 \beta. \\
 }
 for $u_i=(u,c,t), d_i=(d, s, b)$ and $e_i=(e,\mu,\tau)$.
 }
We remark that the parameters defined in Eq. (\ref{cpar}) are invariant under the SM version of the field redefinition (\ref{repar}). 
Having Eq. (\ref{RG-SM-cpar}), a complete set of RG equations below the BSM scale includes the following equations for the running of $\tilde{c}_A \,(A=G, W, B)$.
\bea \label{RG-ctilde-SM}
\frac{ d \tilde{c}_G } { d \ln \mu} &\simeq& - \frac{1}{4\pi^2} y_t^2 C_t + \frac{6\alpha_s^2}{\pi^2} \tilde{c}_G  +\frac{27\alpha_2^2}{16\pi^2} \tilde{c}_W +\frac{11\alpha_1^2}{16\pi^2} \tilde{c}_B\,,  \nonumber \\
\frac{ d \tilde{c}_W } { d \ln \mu} &\simeq& - \frac{3}{16\pi^2} y_t^2 C_t + \frac{9}{2\pi^2} \alpha_s^2\tilde{c}_G  +\frac{27 }{8\pi^2} \alpha_2^2\tilde{c}_W +\frac{3}{8\pi^2}\alpha_1^2 \tilde{c}_B\,,  \nonumber \\
\frac{ d \tilde{c}_B } { d \ln \mu} &\simeq& - \frac{17}{48\pi^2}y_t^2 C_t + \frac{11}{2\pi^2} \alpha_s^2\tilde{c}_G  +\frac{9}{8\pi^2} \alpha_2^2 \tilde{c}_W +\frac{95}{24\pi^2} \alpha_1^2\tilde{c}_B\,. 
\eea


%
{
Below the weak scale ($\mu < m_W$), the top quark, Higgs boson, and $W/Z$ gauge bosons are integrated out. We ignore the corresponding threshold corrections calculated in \cite{2012.12272}
because they are of subleading order giving a correction to our calculation less than 10\%.
At this scale, the relevant axion couplings take the form
 \dis{
 c_\gamma \frac{e^2}{32\pi^2}\frac{a}{f_a} F^{\mu\nu}\tilde F_{\mu\nu}+
\frac{\partial_\mu a}{2 f_a} \Big[ C_{u_i} \bar{u}_i \gamma^\mu \gamma_5 u_i + C_{d_i} \bar{d}_i \gamma^\mu \gamma_5 d_i + C_{e_i} \bar{e}_i \gamma^\mu \gamma_5 e_i \Big],
}
where $F^{\mu\nu}$ is the electromagnetic field strength, and
$u_i, d_i$ and $e_i$ denote the Dirac quarks and leptons with masses lighter than the RG point $\mu$.
Here
\dis{
c_\gamma = c_W + c_B,
}
 and the matching conditions for $C_{u_i}, C_{d_i}$ and $C_{e_i}$ at the weak scale are given in Eq. (\ref{cpar}).
The RG equations for the above axion couplings are given by}
\bea \label{RG-SM-bw}
\frac{d C_{u_i}}{d\ln \mu} & \simeq &   -\frac{\alpha_s^2}{\pi^2} \tilde{c}_G  -\frac{\alpha_{\rm em}^2}{6\pi^2} \tilde{c}_\gamma\,,  \nonumber \\
\frac{dC_{d_i}}{d\ln \mu} & \simeq & -\frac{\alpha_s^2}{\pi^2} \tilde{c}_G   -\frac{\alpha_{\rm em}^2}{24\pi^2} \tilde{c}_\gamma\,,  \nonumber \\
\frac{dC_{e_i}}{d\ln \mu} & \simeq &  -\frac{3\alpha_{\rm em}^2}{8\pi^2} \tilde{c}_\gamma\,,  
\eea
where
\dis{ \label{c-gauge-SM}
\tilde{c}_G(\mu) &= c_G -\sum_q C_q(\mu) \Theta(\mu -m_q), \\
\tilde{c}_\gamma(\mu) &= c_\gamma - 2\sum_f N_c^f Q_f^2 C_f(\mu) \Theta(\mu-m_f),
}
with the unit step function $\Theta(x) = 1$ for $x>0$ and $\Theta(x) =0$ for $x<0$. Here $q$ stands for the quarks $u, d, s, c, b$, and $N_c^f$ and $Q_f$ are the number of colors and the electromagnetic charge of the SM fermion $f$ lighter than the top quark, respectively. 


Finally below the QCD scale ($ \mu \lesssim 1$ GeV),  the relevant axion couplings can be written as 
\bea
\label{c-below-qcd}
\frac{1}{4}g_{a\gamma} a\vec E\cdot \vec B + 
\partial_\mu a \left[
 \frac{g_{ae}}{2m_e} \bar e \gamma^\mu \gamma_5 e+ \frac{g_{an}}{2m_n} \bar n \gamma^\mu \gamma_5 n+\frac{g_{ap}}{2m_p} \bar p \gamma^\mu \gamma_5 p\right],
 \eea
where  the axion-photon coupling $g_{a\gamma}$ includes the
 threshold correction from the axion-pion mixing:
 \bea \label{gag_IR}
 g_{a\gamma} \simeq \frac{\alpha_{\rm em}}{2\pi}\frac{1}{f_a}\Big( c_\gamma -\frac{2}{3}\frac{m_u+4m_d}{m_u+m_d} c_G\Big)\,
 \simeq\,  \frac{\alpha_{\rm em}}{2\pi}\frac{1}{f_a}\Big( c_\gamma-1.92 c_G\Big),
 \eea
the axion-electron coupling $g_{ae}$ evolves down to $\mu = m_e$ as
\dis{
\frac{dg_{ae}}{d\ln \mu} \simeq   -\frac{3\alpha_{\rm em}m_e}{4\pi} g_{a\gamma}
\quad \Big(g_{ae}=\frac{m_e C_e}{f_a}\Big),
}
and
the
axion-nucleon couplings can be computed from the couplings $c_G, C_u(\mu)$ and $C_d(\mu)$ at the matching scale $\mu = 2$ GeV using the lattice data \cite{1511.02867}:
\dis{ 
 g_{ap} &\simeq  \frac{m_p}{f_a}\left( C_u \Delta u + C_d \Delta d -\Big(\frac{m_d}{m_u+m_d}\Delta u +\frac{m_u}{m_u+m_d}\Delta d\Big) c_G\right),\label{Cpcomp} \\
 &\simeq  \frac{m_p}{f_a}\Big(  0.90 \, C_u(2\,{\rm GeV}) -0.38\, C_d(2\,{\rm GeV}) -0.48 \,c_G\Big),}
\dis{
 g_{an} &\simeq  \frac{m_n}{f_a}\left( C_d \Delta u + C_u \Delta d -\Big(\frac{m_u}{m_u+m_d}\Delta u +\frac{m_d}{m_u+m_d}\Delta d\Big) c_G\right), \label{Cncomp} \\
 &\simeq  \frac{m_n}{f_a}\Big(0.90\, C_d(2\,{\rm GeV})-0.38\, C_u(2\,{\rm GeV})-0.04\, c_G\Big), 
 }
with $\Delta u = 0.897(27)$, $\Delta d = -0.376(27)$, and $m_u/m_d = 0.48(3)$ at $\mu = 2$ GeV in $\overline{\rm MS}$ \cite{1511.02867, 2003.01100}.\footnote{Note that the numbers in Eq. (\ref{Cpcomp}) and Eq. (\ref{Cncomp}) are a bit different from those in Eq. 50 of Ref. \cite{1511.02867}. This is because here our $C_u$ and $C_d$ are not UV boundary parameters around $\mu = f_a$ but the renormalized parameters at $\mu = 2$ GeV. If we express $C_u(2\, \textrm{GeV})$ and $C_d(2\, \textrm{GeV})$ in terms of the UV parameters, for instance, using Eq. (\ref{SMnum}) in the following analysis, we get better agreement with those numbers in Ref. \cite{1511.02867} except the coefficient of $C_t(f_a)$. Our coefficient of $C_t(f_a)$ from Eq. (\ref{SMnum}) is significantly different from the one in Ref. \cite{1511.02867}, because Ref. \cite{1511.02867} does not take into account the renormalization by Yukawa interactions.}

\subsection{Semi-analytic solution and numerical results}

We now discuss the solution to the RG equations of the previous subsection. We are concerned with the RG corrections to the low energy couplings $C_u, C_d$ and $C_e$. We will provide semi-analytic formulas for them as well as showing fully numerical results.

For a small $\tan \beta (\lesssim 50)$, as a good approximation, one can keep only the top Yukawa coupling $\tilde{y}_t$ among the three Yukawa couplings $\tilde{y}_{t, b, \tau}$ in the RG equations (\ref{RG-npar}). By integrating the RG equations Eq. (\ref{RG-npar}), Eq. (\ref{RG-SM-cpar}), and Eq. (\ref{RG-SM-bw}) using this approximation, we get for a renormalization scale $\mu < m_W$ 
\dis{\label{axion-uquark}
C_u& (\mu) \simeq C_u (f_a) + I_t (m_{\rm BSM}, f_a) \sin^2 \beta + I_t^{\rm SM} (m_t, m_{\rm BSM})  \\
&+ \frac{2}{3\pi^2} \int_{m_{\rm BSM}}^{f_a}\frac{d\mu'}{\mu'}  \left( \alpha_s^2 (\mu') \tilde{c}_G(\mu')
+ \frac{9\sin^2 \beta}{16} \alpha_2^2(\mu') \tilde{c}_W(\mu') +\frac{(4+9\sin^2 \beta)}{48} \alpha_1^2 (\mu') \tilde{c}_B(\mu') \right), \\
&+ \frac{1}{\pi^2} \int_{m_W}^{m_{\rm BSM}}\frac{d\mu'}{\mu'}  \left(\alpha_s^2 (\mu') \tilde{c}_G(\mu')
+ \frac{9}{32} \alpha_2^2(\mu') \tilde{c}_W(\mu') +\frac{17}{96} \alpha_1^2 (\mu') \tilde{c}_B(\mu') \right), \\
&+\frac{1}{\pi^2} \int_{\mu}^{m_W}\frac{d\mu'}{\mu'}  \left(\alpha_s^2 (\mu') \tilde{c}_G(\mu')
+ \frac{1}{6} \alpha_{\rm em}^2(\mu') \tilde{c}_\gamma(\mu') \right), 
}
\dis{\label{axion-dquark}
C_d& (\mu) \simeq C_d (f_a) - I_t (m_{\rm BSM}, f_a) \sin^2 \beta - I_t^{\rm SM} (m_t, m_{\rm BSM})  \\
&+ \frac{2}{3\pi^2} \int_{m_{\rm BSM}}^{f_a}\frac{d\mu'}{\mu'}  \left( \alpha_s^2 (\mu') \tilde{c}_G(\mu')
+ \frac{9\cos^2 \beta}{16} \alpha_2^2(\mu') \tilde{c}_W(\mu') +\frac{(-2+9\cos^2 \beta)}{48} \alpha_1^2 (\mu') \tilde{c}_B(\mu') \right), \\
&+ \frac{1}{\pi^2} \int_{m_W}^{m_{\rm BSM}}\frac{d\mu'}{\mu'}  \left(\alpha_s^2 (\mu') \tilde{c}_G(\mu')
+ \frac{9}{32} \alpha_2^2(\mu') \tilde{c}_W(\mu') +\frac{5}{96} \alpha_1^2 (\mu') \tilde{c}_B(\mu') \right), \\
&+\frac{1}{\pi^2} \int_{\mu}^{m_W}\frac{d\mu'}{\mu'}  \left(\alpha_s^2 (\mu') \tilde{c}_G(\mu')
+ \frac{1}{24} \alpha_{\rm em}^2(\mu') \tilde{c}_\gamma(\mu') \right), 
}
\dis{
\label{axion-electron}
C_e (\mu) \simeq&~ C_e (f_a) - I_t (m_{\rm BSM}, f_a) \sin^2 \beta - I_t^{\rm SM} (m_t, m_{\rm BSM})  \\
&+ \frac{2}{3\pi^2} \int_{m_{\rm BSM}}^{f_a}\frac{d\mu'}{\mu'}  \left( \frac{9\cos^2 \beta}{16} \alpha_2^2(\mu') \tilde{c}_W(\mu') +\frac{3(2+\cos^2 \beta)}{16} \alpha_1^2 (\mu') \tilde{c}_B(\mu') \right), \\
&+ \frac{1}{\pi^2} \int_{m_W}^{m_{\rm BSM}}\frac{d\mu'}{\mu'}  \left( \frac{9}{32} \alpha_2^2(\mu') \tilde{c}_W(\mu') +\frac{15}{32} \alpha_1^2 (\mu') \tilde{c}_B(\mu') \right), \\
&+\frac{1}{\pi^2} \int_{\mu}^{m_W}\frac{d\mu'}{\mu'}  \frac{3}{8} \alpha_{\rm em}^2(\mu') \tilde{c}_\gamma(\mu') , 
}
where $C_\Psi(f_a) \equiv C_\Psi^0$ as defined in Eq. (\ref{tree_coupling}), and $I_t(m_{\rm BSM}, f_a)$ and $I_t^{\rm SM}(m_t, m_{\rm BSM})$ are defined as
\bea
\label{axion-tquark}
I_t(m_{\rm BSM}, f_a) &\equiv& -\frac{3}{8\pi^2} \int_{m_{\rm BSM}}^{f_a} \frac{d \mu'}{\mu'} \tilde{y}_t^2 (\mu') n_t(\mu'), \\
I_t^{\rm SM}(m_t, m_{\rm BSM}) &\equiv& -\frac{3}{8\pi^2} \int_{m_t}^{m_{\rm BSM}} \frac{d \mu'}{\mu'} y_t^2 (\mu') C_t(\mu').
\eea
In the above integrations one can ignore the running of the parameters $\tilde{c}_G(\mu), \tilde{c}_W(\mu)$ and $\tilde{c}_B(\mu)$ as this effect is of subleading order, while one has to take into account the heavy particle threshold corrections as in Eq. (\ref{c-gauge-SM}). 

The above formulas show that the RG corrections to $C_{u, d, e}$ are mainly determined by the running parameters $n_t, C_t, \tilde{c}_{G, W, B}$. 
To get a more instructive form for $I_t$ and $I_t^{SM}$  from the top Yukawa interaction, one may integrate the one-loop RG equations for $\alpha_s(\mu)$ and the top Yukawa coupling while solving the first-order linear differential equations for $n_t(\mu)$ and $C_t(\mu)$ in Eq. (\ref{RG-npar}) and Eq. (\ref{RG-SM-cpar}). We then find
\dis{ \label{It_BSM}
I_t&(m_{\rm BSM}, f_a) \simeq -\frac12 \left(1-R_t(m_{\rm BSM}, f_a)\right) n_t (f_a) \\ 
&-\frac{1}{3\pi^2} \int_{m_{\rm BSM}}^{f_a} \frac{d \mu'}{\mu'}  \left(1-R_t(m_{\rm BSM}, \mu')\right)\left(\alpha_s^2(\mu') \tilde{c}_G(\mu') + \frac{9}{16} \alpha_2^2(\mu') \tilde{c}_W(\mu')
+\frac{13}{48} \alpha_1^2(\mu')\tilde{c}_B(\mu') \right),
}
\dis{ \label{It_SM}
I_t^{\rm SM}&(m_t, m_{\rm BSM}) \simeq -\frac23 \left(1-R_t^{\rm SM}(m_t, m_{\rm BSM})\right) C_t (m_{\rm BSM}) \\ 
&-\frac{2}{3\pi^2} \int_{m_t}^{m_{\rm BSM}} \frac{d \mu'}{\mu'}  \left(1-R_t^{\rm SM}(m_t, \mu')\right)\left(\alpha_s^2(\mu') \tilde{c}_G(\mu') + \frac{9}{32} \alpha_2^2(\mu') \tilde{c}_W(\mu')
+\frac{17}{96} \alpha_1^2(\mu')\tilde{c}_B(\mu') \right),
}
with
\dis{
R_t(\mu, \mu') &\equiv \exp\left[-24  \int_{\mu}^{\mu'} \frac{d\mu''}{\mu''} \frac{\tilde{y}_t^2(\mu'')}{32\pi^2}\right] \simeq \left[1 - \frac{18}{7} \frac{\alpha_t(\mu')}{\alpha_s(\mu')} \left(1-\left(\frac{\alpha_s(\mu)}{\alpha_s(\mu')} \right)^{7/9} \right)\right]^{-1},
}
\dis{
R_t^{\rm SM} (\mu, \mu') &\equiv \exp\left[-18  \int_{\mu}^{\mu'} \frac{d\mu''}{\mu''} \frac{y_t^2(\mu'')}{32\pi^2}\right] \simeq \left[1 - \frac{9}{2} \frac{\alpha_t(\mu')}{\alpha_s(\mu')} \left(1-\left(\frac{\alpha_s(\mu)}{\alpha_s(\mu')} \right)^{1/7} \right)\right]^{-1},
}
where $\alpha_t (\mu) \equiv \tilde{y}_t^2(\mu)/4\pi$ for $\mu > m_{\rm BSM}$ and $\alpha_t (\mu) \equiv y_t^2(\mu)/4\pi$ for $\mu < m_{\rm BSM}$. Eq. (\ref{It_BSM}) and Eq. (\ref{It_SM}) explicitly show potentially important contributions from $\tilde{c}_{G, W, B}$ through the top Yukawa interaction even when $n_t(f_a)$ and $C_t(m_{\rm BSM})$ vanish. As we will see below, they give rise to an important radiative correction to $C_e$ for KSVZ-like axions.

\begin{figure}[th]  
  \includegraphics[width=1.0 \textwidth]{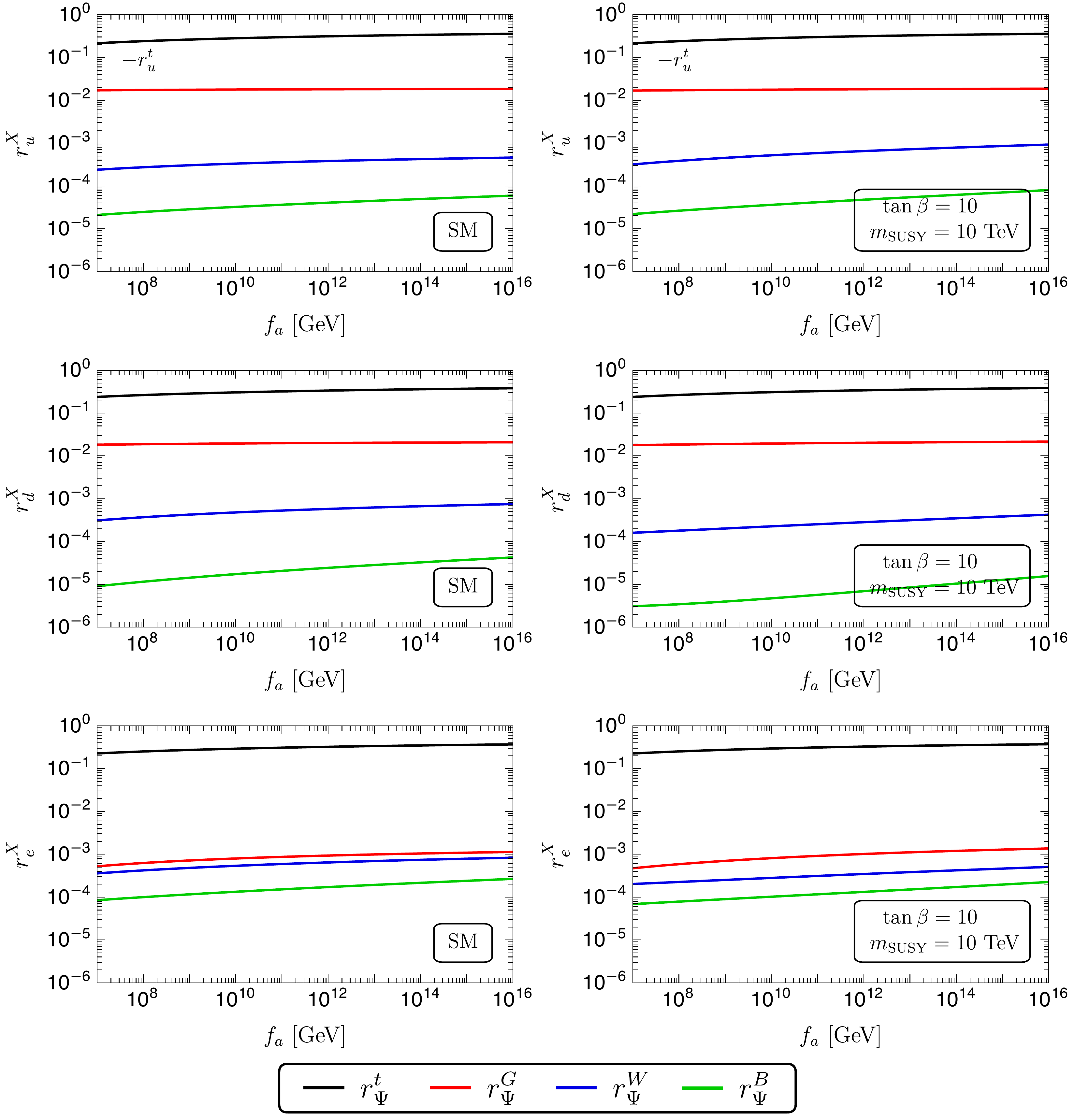} 
  \caption{Radiative corrections to the axion couplings to the light quarks and electron from  $\left\{C_t(f_a), \tilde{c}_{G}(f_a), \tilde{c}_W(f_a), \tilde{c}_B(f_a)\right\}$ for the SM or from $\left\{n_t(f_a), \tilde{c}_{G}(f_a), \tilde{c}_W(f_a), \tilde{c}_B(f_a)\right\}$ for the MSSM :  $\Delta C_\Psi=  r_\Psi^t \left[ n_t(f_a) ~\textrm{or} ~C_t(f_a) \right] + \sum_{A=G, W, B} r_{\Psi}^A \tilde{c}_A (f_a)$.  
 For the MSSM, we take the parameters $\tan \beta = 10$ and $m_{\rm SUSY}=10$ TeV.  }
 \label{numeric_sol}
 \end{figure}

{ Eqs. (\ref{axion-uquark}), (\ref{axion-dquark}), (\ref{axion-electron}) and (\ref{axion-tquark}) show that the radiative corrections to $C_\Psi$ $(\Psi =u, d, e)$ induced by the top Yukawa interaction and the SM gauge interactions originate mainly from
the four UV parameters $\{n_t(f_a), \tilde{c}_{G}(f_a), \tilde{c}_W(f_a), \tilde{c}_B(f_a) \}$ in the MSSM case or from $\{C_t(f_a), \tilde{c}_{G}(f_a), \tilde{c}_W(f_a), \tilde{c}_B(f_a)\}$ in the SM case. Then one can parameterize 
the low energy axion couplings to the light quarks and electron  as
\bea
C_u(2\, {\rm GeV})&=& C_u(f_a)+\Delta C_u,\nonumber \\
C_d(2\, {\rm GeV})&=& C_d(f_a)+\Delta C_d, \nonumber \\
C_e(m_e)&=&C_e(f_a)+\Delta C_e,\eea
where $C_\Psi(f_a)$ ($\Psi=u,d,e$) are identified as the tree-level values $C^0_\Psi$ and 
\bea
&\mbox{SM}:&\quad 
\Delta C_\Psi = r_{\Psi}^t C_t(f_a)  + r_{\Psi}^G \tilde{c}_G (f_a) + r_{\Psi}^W \tilde{c}_W (f_a) +  r_{\Psi}^B \tilde{c}_B (f_a),\nonumber \\
&\mbox{MSSM}:&\quad 
\Delta C_\Psi = r_{\Psi}^t  n_t(f_a) + r_{\Psi}^G \tilde{c}_G (f_a) + r_{\Psi}^W \tilde{c}_W (f_a) +  r_{\Psi}^B \tilde{c}_B (f_a). 
\eea 

To compute the coefficients $r_\Psi^{X}\, (X=t, G, W, B)$ in this parameterization, we solve the RG equations in a fully numerical way  with the running SM gauge couplings and top Yukawa coupling at two-loop order.
The resulting  $r_\Psi^X$ are depicted  in
 Fig. \ref{numeric_sol} for the SM case (left panel) and the MSSM case (right panel). For the MSSM, we choose $\tan \beta = 10$ and  the SUSY particle masses $m_{\rm SUSY} = 10$ TeV. 
Our results show that
\bea
r^t_{u,d,e}\sim\, {\rm few}\times 10^{-1}, \quad r^G_{u,d}\sim 10^{-2}, \quad r^G_e\sim \, {\rm few}\times 10^{-4}-10^{-3},\nonumber \\
r^W_{u,d,e}\sim \,{\rm few}\times 10^{-4}-10^{-3}, \quad r^B_{u,d}\sim \,{\rm few}\times 10^{-5}-10^{-4}, \quad r^B_e\sim 10^{-4}
 \eea
 for $10^7\, {\rm GeV}\lesssim f_a \lesssim 10^{16}\, {\rm GeV}$.
For example, for $f_a=10^{10}$ GeV in the SM case, we find
\dis{ \label{SMnum}
C_u(2\,{\rm GeV}) &\simeq C_u(f_a) -0.28\, C_t(f_a) + \left[ 17.8 \,\tilde{c}_G(f_a) +0.33\,\tilde{c}_W(f_a)+0.032\,\tilde{c}_B(f_a)  \right]\times 10^{-3}\,, \\
C_d(2\,{\rm GeV}) &\simeq C_d(f_a) +0.30 \,C_t(f_a) + \left[ 19.5 \,\tilde{c}_G(f_a) +0.48\,\tilde{c}_W(f_a)+0.017\,\tilde{c}_B(f_a)  \right] \times 10^{-3}\,, \\
C_e(m_e) &\simeq C_e(f_a) +0.29 \,C_t(f_a) + \left[ 0.80 \,\tilde{c}_G(f_a) +0.54\,\tilde{c}_W(f_a)+0.13\,\tilde{c}_B(f_a)  \right] \times 10^{-3}\,,
}
and  for $f_a=10^{10}$ GeV, $m_{\rm SUSY} =10$ TeV and $\tan \beta = 10$ in the MSSM case,  
\dis{ \label{MSSMnum}
C_u(2\,{\rm GeV}) &\simeq C_u(f_a) -0.28 \,n_t(f_a) + \left[ 17.7 \,\tilde{c}_G(f_a) +0.52\,\tilde{c}_W(f_a)+0.036\,\tilde{c}_B(f_a)  \right]\times 10^{-3}\,,  \\
C_d(2\,{\rm GeV}) &\simeq C_d(f_a) +0.31 \,n_t(f_a) + \left[ 19.4 \,\tilde{c}_G(f_a) +0.23\,\tilde{c}_W(f_a)+0.0047\,\tilde{c}_B(f_a)  \right] \times 10^{-3}\,, \\
C_e(m_e) &\simeq C_e(f_a) +0.29 \,n_t(f_a) + \left[ 0.81 \,\tilde{c}_G(f_a) +0.28\,\tilde{c}_W(f_a)+0.10\,\tilde{c}_B(f_a)  \right] \times 10^{-3}\,. 
}
Here the difference in the value of $r^W_\Psi$
between the SM case and the MSSM case is mainly due to
a different
running of the gauge couplings and top Yukawa coupling.
}
\begin{figure}  
\centering
\includegraphics[scale=0.25]{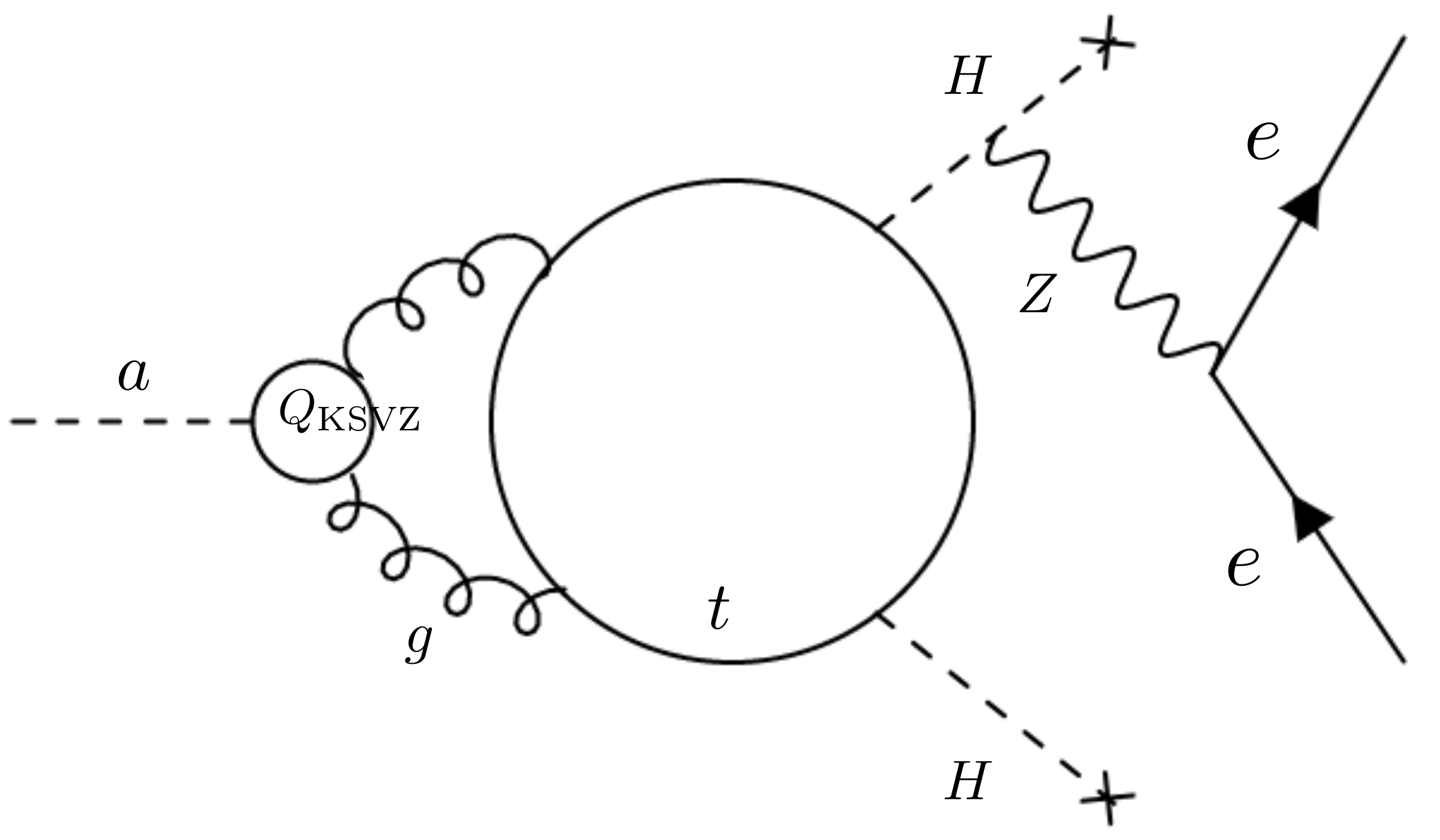}  \hspace{0.2cm}
\includegraphics[scale=0.3]{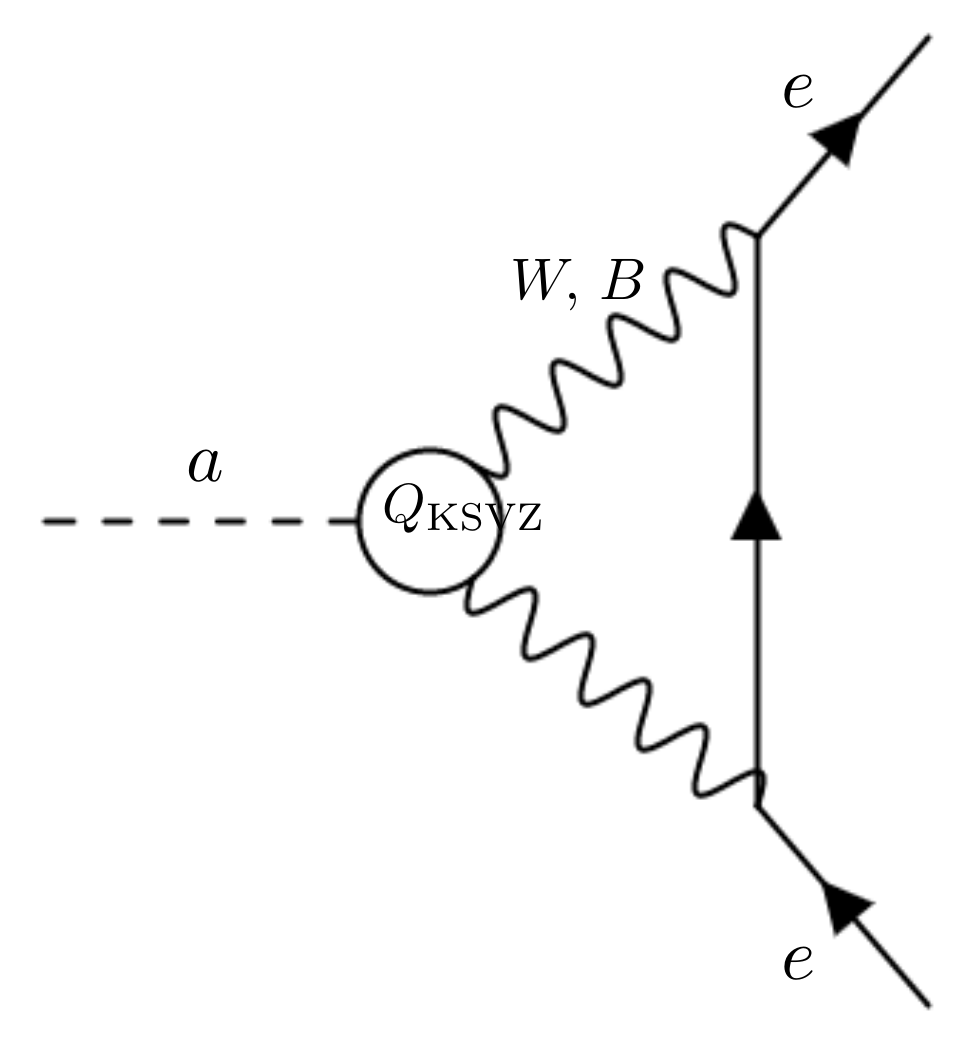} \hspace{0.1cm}
\includegraphics[scale=0.3]{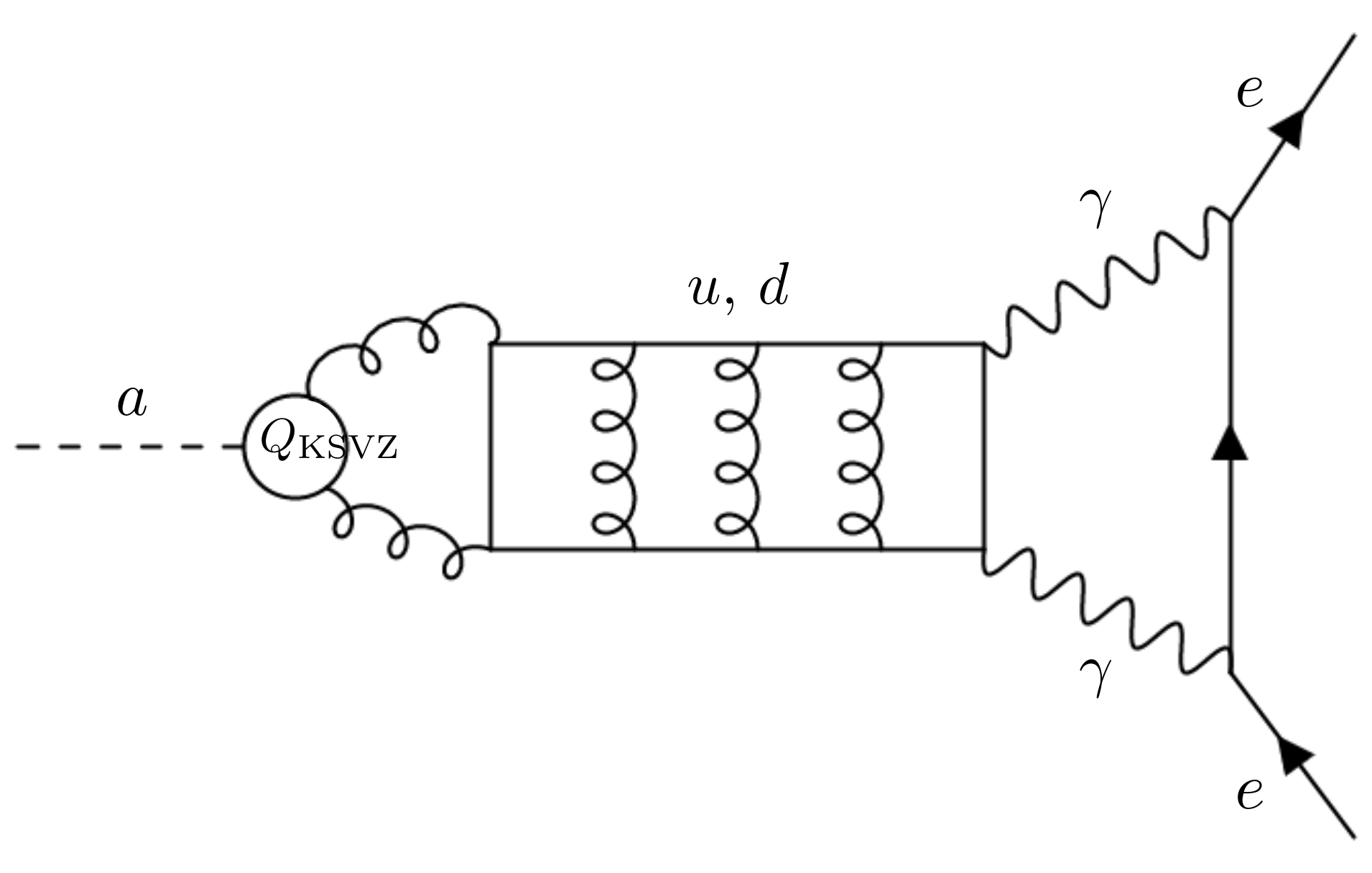}
\caption{Diagrams for the radiative corrections to the axion-electron coupling $C_e$ in KSVZ-like models.}
\label{fig:KSVZ}
\end{figure}

{
A particularly interesting aspect of our results  is that the axion-electron coupling $C_e$ receives a large radiative correction from the UV parameter $\tilde{c}_G(f_a)$, which was first shown in \cite{2012.12272} for the SM case to our knowledge. This has an important implication for KSVZ-like QCD axions.
For KSVZ-like axions, $n_t(f_a)=C_t(f_a)=0$ so that there is no one-loop level radiative correction from the Yukawa interactions. 
In the previous literature, the leading contribution to $C_e$ in KSVZ axion model was thought to be from non-zero $c_W$ or $c_B$ as in the second diagram of Fig. \ref{fig:KSVZ}, or from the axion-pion mixing below the QCD scale  as in the third diagram of Fig. \ref{fig:KSVZ} for a minimal KSVZ model 
 with $c_W=c_B=0$
\cite{Srednicki:1985xd, hep-ph/9306216}. 
But the results Eq. (\ref{SMnum}) and Eq. (\ref{MSSMnum}) indicate that there is a larger contribution from the axion-gluon coupling $c_G$. 
This radiative correction is originated from the first diagram of Fig. \ref{fig:KSVZ} involving the exotic heavy quark (${\cal Q},{\cal Q}^c$), gluons, top quark and Higgs
doublet at three-loop level. This process is encoded in $I_t$ and $I_t^{\rm SM}$ (Eq. (\ref{It_BSM}) and Eq. (\ref{It_SM})) of Eq. (\ref{axion-electron}). Although it is a three-loop process, it beats the contributions from the second and third diagrams in Fig. \ref{fig:KSVZ} because $\alpha_s^2 \gg \alpha_{2, 1}^2$ and $y_t \sim {\cal O}(1)$.  
For instance,   for KSVZ-like axions with
 $m_{\cal Q} = 10^{10}$ GeV, we find
\bea
C_e(m_e) &\simeq&  \Big[ 0.84\,c_G-0.03\, c_G+0.28\,c_W+0.10\,c_B  \Big] \times 10^{-3}\,~  (\textrm{KSVZ with MSSM}) \nonumber \\
C_e(m_e) &\simeq&  \Big[ 0.83\,c_G-0.03\, c_G+0.54\,c_W+0.13\,c_B  \Big] \times 10^{-3}\,~   (\textrm{KSVZ with SM}),
\eea
where the first term in the square bracket  denotes the contribution from the first diagram in Fig. \ref{fig:KSVZ}, the second  is the contribution from the axion-pion mixing represented by the last diagram in  Fig. \ref{fig:KSVZ}, and the last two terms are the contributions from the two-loop diagrams involving the electroweak gauge bosons.
 }

\section{Distinguishing the axions by low energy precision physics} \label{sec:PP}
 

Based on the results of the previous sections, we are going to discuss the possibility to distinguish the different classes of axion models experimentally. As discussed in section \ref{sec:UV}, one can consider three classes of axion models: KSVZ-like, DFSZ-like, and string-theoretic axions. 
They may show different patterns of the low energy axion couplings to the nucleons and electron after properly taking into account the relevant radiative corrections.
{To include string-theoretic axions in a consistent manner, in the following we discuss the coupling patterns in the MSSM framework which assumes $N=1$ SUSY at scales around $f_a$. 
}



Let us first consider the approximate expression of 
the low energy axion couplings defined in Eq. (\ref{c-below-qcd})
 for a specific parameter point with $f_a=10^{10}$ GeV, $m_{\rm SUSY} = 10$ TeV and $\tan \beta=10$. 
Applying the numerical result in Eq. (\ref{MSSMnum}) to Eq. (\ref{Cpcomp}) and Eq. (\ref{Cncomp}), we find
\bea
 g_{a p} &\simeq& \frac{m_p}{f_a} \Big[-0.48\, c_G +0.90\, C_u^0 -0.38\, C_d^0 \Big.  \nonumber\\ 
 &&\Big. \quad\quad~-0.37\, n_t(f_a) +  \Big(8.6 \tilde{c}_G(f_a) + 0.38 \tilde{c}_W(f_a) +  0.031 \tilde{c}_B(f_a)\Big)\times 10^{-3} \Big], \nonumber\\
 g_{a n}  &\simeq& \frac{m_n}{f_a} \Big[-0.037\, c_G -0.38\, C_u^0 +0.90\, C_d^0 \Big. \nonumber\\ 
 &&\Big. \quad\quad~+0.38\, n_t(f_a) +  \Big(11 \tilde{c}_G(f_a) + 0.0078 \tilde{c}_W(f_a) -  0.0094 \tilde{c}_B(f_a)\Big)\times 10^{-3} \Big], \nonumber\\
 g_{a e}  &\simeq& \frac{m_e}{f_a} \Big[ \,C_e^0 +0.29\, n_t(f_a) +  \Big(0.81 \tilde{c}_G(f_a) + 0.28 \tilde{c}_W(f_a) +  0.10 \tilde{c}_B(f_a)\Big)\times 10^{-3} \Big], 
\eea
where $C_u^0, C_d^0$ and $C_e^0$ are the tree level couplings as defined in Eq. (\ref{tree_coupling}), and  the contributions from $n_t(f_a)$ and $\tilde{c}_A(f_a) \,(A= G, W, B)$ represent the radiative corrections. On the other hand,
the axion-photon coupling $g_{a \gamma}$ at low energies below the QCD scale is given by Eq. (\ref{gag_IR}),
\dis{
 g_{a \gamma} \simeq \frac{\alpha_{\rm em}}{2\pi} \frac{1}{f_a} \Big(c_W + c_B -1.92 c_G\Big).
 }

As we have defined in section \ref{sec:UV}, we refer to DFSZ-like axions as generic field-theoretic axions with $C_\Psi^0 = {\cal O}(1)~(\Psi=u, d, e)$. Thus for SUSY DFSZ-like axions with $f_a=10^{10}$ GeV, $m_{\rm SUSY} = 10$ TeV and $\tan \beta=10$,
\bea \label{g_DFSZ}
 g_{a p}^{\rm DFSZ}   &\simeq& \frac{m_p}{f_a} \left[-0.48\, c_G +0.90\, C_u^0 -0.38\, C_d^0 -0.37\, n_t(f_a)  \right] \sim  \frac{m_p}{f_a} {\cal O}(1), \nonumber\\
 g_{a n}^{\rm DFSZ}   &\simeq& \frac{m_n}{f_a} \left[-0.037\, c_G -0.38\, C_u^0 +0.90\, C_d^0 +0.38\, n_t(f_a)  \right] \sim  \frac{m_n}{f_a} {\cal O}(1), \nonumber\\
 g_{a e}^{\rm DFSZ}   &\simeq& \frac{m_e}{f_a} \left[ \,C_e^0 +0.29\, n_t(f_a) \right] \sim  \frac{m_e}{f_a} {\cal O}(1).
\eea
Here we note $n_t(f_a)=0$ for the standard DFSZ models of Eq. (\ref{DFSZpar}) and Eq. (\ref{SUSYDFSZpar}), while it can be non-vanishing for more generic DFSZ-like axions \cite{2101.03173}.
In fact, for the standard DFSZ axions, $n_t(f_a) = \tilde{c}_{A}(f_a) =0 \, (A=G, W, B)$, and the RG running is present only below the BSM scale after the BSM Higgs doublet or Higgsinos are integrated out giving rise to a non-zero $C_t(m_{\rm BSM})$ or $\tilde{c}_{W, B}(m_{\rm BSM})$, while the running by $\tilde{c}_G$ is present only after the top quark is integrated out. Thus the RG running effect is mostly negligible for the standard DFSZ models, while there could be a sizable RG correction from the top Yukawa coupling for more generic DFSZ-like axions having a non-zero $n_t(f_a)$.

On the other hand, KSVZ-like axions are defined as generic field-theoretic axions having vanishing tree level couplings to the SM particles and Higgs doublets, i.e. $C_u^0=C_d^0= n_t(f_a) =0$ {(or more generically $\lesssim {\cal O}\big((g_{\rm GUT}^2/8\pi^2)^2\big)$} and ~$\tilde{c}_A(f_a) = c_A ~ (A=G, W, B)$. Thus for SUSY KSVZ-like axions with $f_a=10^{10}$ GeV, $m_{\rm SUSY} = 10$ TeV and $\tan \beta=10$,
\bea \label{g_KSVZ}
 g_{a p}^{\rm KSVZ}   &\simeq& \frac{m_p}{f_a} \left[ -0.48c_G + (0.38 c_W + 0.031 c_B)\times 10^{-3}  \right], \nonumber\\
 g_{a n}^{\rm KSVZ}  &\simeq& \frac{m_n}{f_a} \left[-0.026c_G +  (0.78 c_W  - 0.94 c_B)\times 10^{-5}  \right], \nonumber\\
 g_{a e}^{\rm KSVZ}  &\simeq& \frac{m_e}{f_a}  \,(0.81c_G  + 0.28 c_W + 0.10 c_B) \times 10^{-3}.
 \eea
In fact here it is assumed that the mass of the exotic heavy quark ${\cal Q}+{\cal Q}^c$ in the model (\ref{KSVZmod}), i.e. $m_{\cal Q} = y_{\cal Q} f_a /\sqrt{2}$, is around $f_a$. If $m_{\cal Q} \ll f_a$ with $y_{\cal Q} \ll 1$, $\tilde{c}_A(f_a) = c_A - c_{\cal Q} - c_{{\cal Q}^c}  =0$, and the RG running of axion couplings starts from the scale $\mu = m_{\cal Q} (\ll f_a)$ at which ${\cal Q}+{\cal Q}^c$ is integrated out, yielding $\tilde{c}_A(m_{\cal Q})=c_A$. For this case, the coefficients in Eq. (\ref{g_KSVZ}) depends on $m_{\cal Q}$ rather than $f_a$. In the following analysis, we will also consider the case $m_{\cal Q} \ll f_a$.
 
String-theoretic axions have a distinctive feature that 
their couplings to matter fields are suppressed by $1/\tau$ where $\tau$ denotes the
moduli partner whose vacuum expectation value can be identified as the Euclidean action
of the brane instanton coupled to string-theoretic axion.  As a consequence,
$C_\Psi^0$ ($\Psi=u,d,e$) are numerically
of ${\cal O}(g_{\rm GUT}^2/8\pi^2)$  for light string-theoretic axions which couple to the gluons or the photon, including a QCD axion and an ultralight ALP coupled to the photon (See Eqs. (\ref{estimate}) and (\ref{estimate_omega})). 
As discussed in section \ref{sec:UV}, one can 
parameterize the string-theoretic axion couplings to matter fields at $\mu\sim f_a$ as
\bea
c_\phi(f_a)=\omega_\phi \frac{g_{\rm GUT}^2}{16\pi^2}, \quad c_\psi =\omega_\psi \frac{g_{\rm GUT}^2}{16\pi^2},
\eea
where the model-dependent coefficients $\omega_\phi$ and $\omega_\psi$ are expected to be of order unity.
We then 
find for string-theoretic axions with $f_a=10^{10}$ GeV, $m_{\rm SUSY} = 10$ TeV and $\tan \beta=10$,
\bea \label{g_string}
 g_{a p}^{\rm string}  &\simeq& \frac{m_p}{f_a} \Big[ g_{\rm GUT}^2 \Big(0.56(\omega_{u^c_1} + \omega_{H_u}) +0.32 \omega_{Q_1} -0.24 (\omega_{Q_3} + \omega_{u^c_3} + \omega_{d^c_1}) \Big)\times 10^{-2} \Big.  \nonumber\\ 
 &&\Big. \quad\quad~ -0.48c_G + (0.38 c_W + 0.031 c_B)\times 10^{-3} \Big], \nonumber\\
 g_{a n}^{\rm string}  &\simeq& \frac{m_n}{f_a} \Big[ g_{\rm GUT}^2 \Big(0.56(\omega_{d^c_1} - \omega_{H_u}) +0.32 \omega_{Q_1} +0.24 (\omega_{Q_3} + \omega_{u^c_3} - \omega_{u^c_1}) \Big)\times 10^{-2} \Big. \nonumber\\ 
 &&\Big. \quad\quad~ -0.026c_G +  (0.78 c_W  - 0.94 c_B)\times 10^{-5}  \Big], \nonumber\\
 g_{a e}^{\rm string}  &\simeq& \frac{m_e}{f_a} \Big[ g_{\rm GUT}^2 \Big(0.63(\omega_{e^c_1} + \omega_{L_1}) -0.44 \omega_{H_u} +0.19 (\omega_{Q_3} + \omega_{u^c_3}) \Big)\times 10^{-2} \Big. \nonumber\\
 &&\Big. \quad\quad~ +(0.81c_G  + 0.28 c_W + 0.10 c_B) \times 10^{-3} \Big].
\eea

\begin{figure}[h]  
 \begin{tabular}{c}
    \includegraphics[width=0.5 \textwidth]{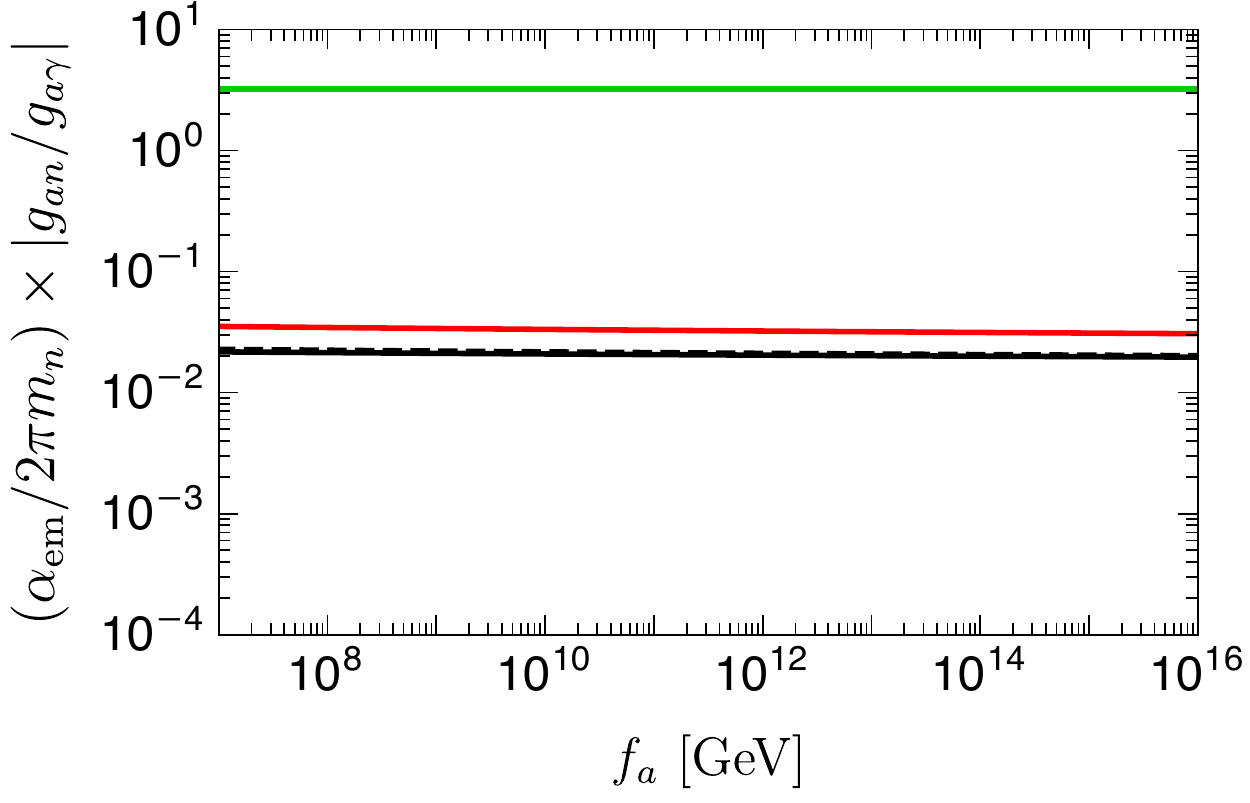} 
   \includegraphics[width=0.5 \textwidth]{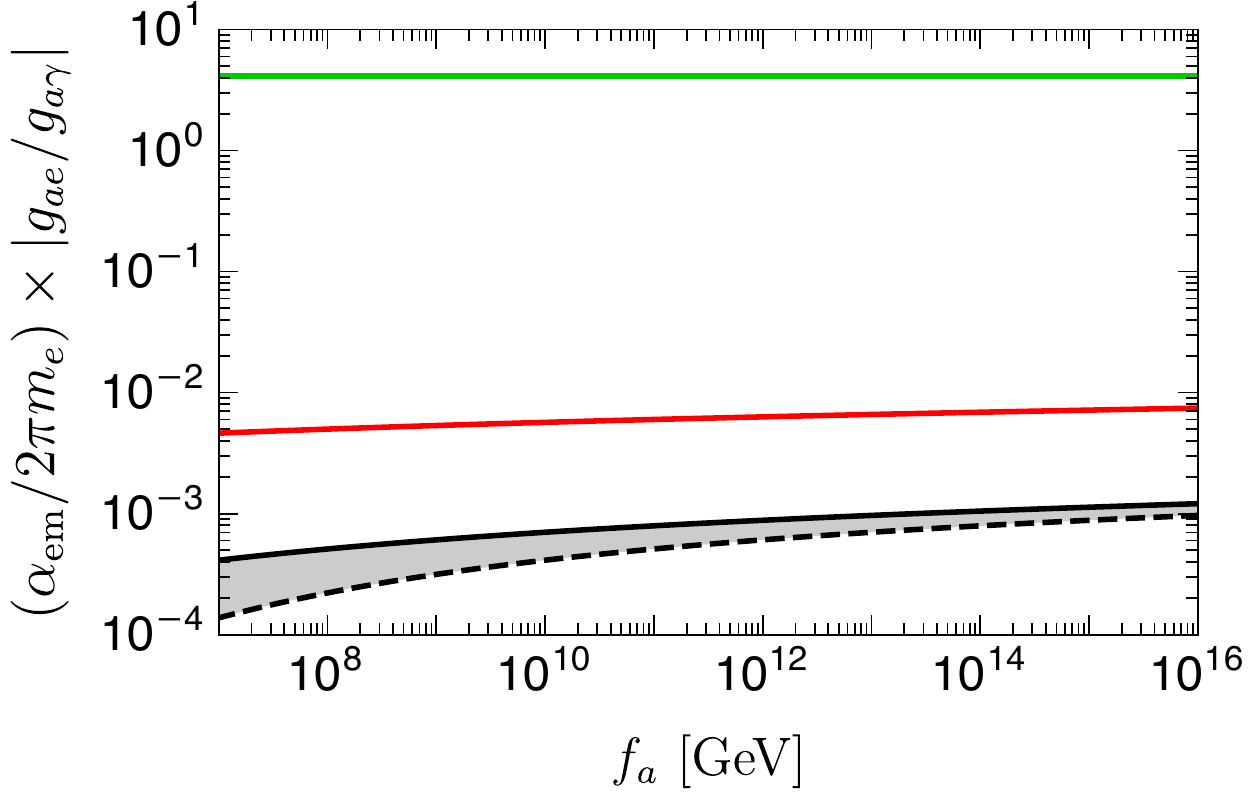} 
  \end{tabular} 
  \caption{ The predicted ratios of the axion couplings from the minimal QCD axion models  
  within the MSSM framework (green: DFSZ-like axions, black: KSVZ-like axions, red: string-theoretic axions). For KSVZ-like axions, the heavy exotic quark is assumed to have a mass between $10^{-3} f_a$ (dashed black) and $f_a$ (solid black).  For the MSSM parameters, we take $\tan \beta = 10$ and $m_{\rm SUSY}=10$ TeV. }
 \label{ratio_plot} \label{QCDaxionplot}
 \end{figure}

From the above results,
one can see that the couplings show qualitatively different patterns depending on whether $c_G \neq 0$ (as for the QCD axion) or $c_G =0$. Therefore let us consider the two cases separately. 

{
For the case with $c_G \neq 0$ that includes the QCD axion, all of the three classes of axions have a similar value of the coupling ratio $g_{ap}/g_{a\gamma}$.  On the other hand,  the  coupling ratio  $g_{an}/g_{a\gamma}$ for KSVZ-like QCD axions and string-theoretic QCD axions
is an order of magnitude smaller  than that of DFSZ-like QCD axions. So one can discriminate between those two classes of models with an experimental measurement of $g_{an}/g_{a\gamma}$. To completely distinguish among the three classes of QCD axions, one needs to measure another coupling ratio $g_{ae}/g_{a\gamma}$ as it is predicted to be of different orders of magnitude depending on the classes of axions. In Fig. \ref{QCDaxionplot}, we numerically plot the predicted coupling ratios $g_{an}/g_{a\gamma}$ and $g_{ae}/g_{a\gamma}$ for each class of QCD axions  over  $f_a$ between $10^7$ GeV and $10^{16}$ GeV.
For DFSZ-like QCD axion, we use the model specified in Eq. (\ref{DFSZtree}) and Eq. (\ref{SUSYDFSZpar}) as a benchmark model. For KSVZ-like QCD  axion, we use the model in Eq. (\ref{KSVZmod}) with an $SU(2)_L$-singlet exotic heavy quark ${\cal Q}$ carrying a $U(1)_Y$ hypercharge $Y({\cal Q})=-1/3$, which implies $c_G=1,\, c_W=0,\, c_B=2/3$. Moreover, we assume the exotic heavy quark has a mass $m_{\cal Q}$ between $10^{-3} f_a$ and $f_a$, which corresponds to the scale at which the axion coupling starts to run for this class of models.
 For a benchmark model of string-theoretic QCD axions,
we use the large volume scenario model in Eq. (\ref{lvs}) with
a universal scaling weight $\omega_I=1/2$ as in Eq. (\ref{sw1/2}), and also
 $g_{\rm GUT}^2 \simeq 1/2$ with the GUT relation $c_G = c_W = (3/5) c_B$. }


\begin{figure}[th]  
 \begin{tabular}{c}
      \includegraphics[width=0.5 \textwidth]{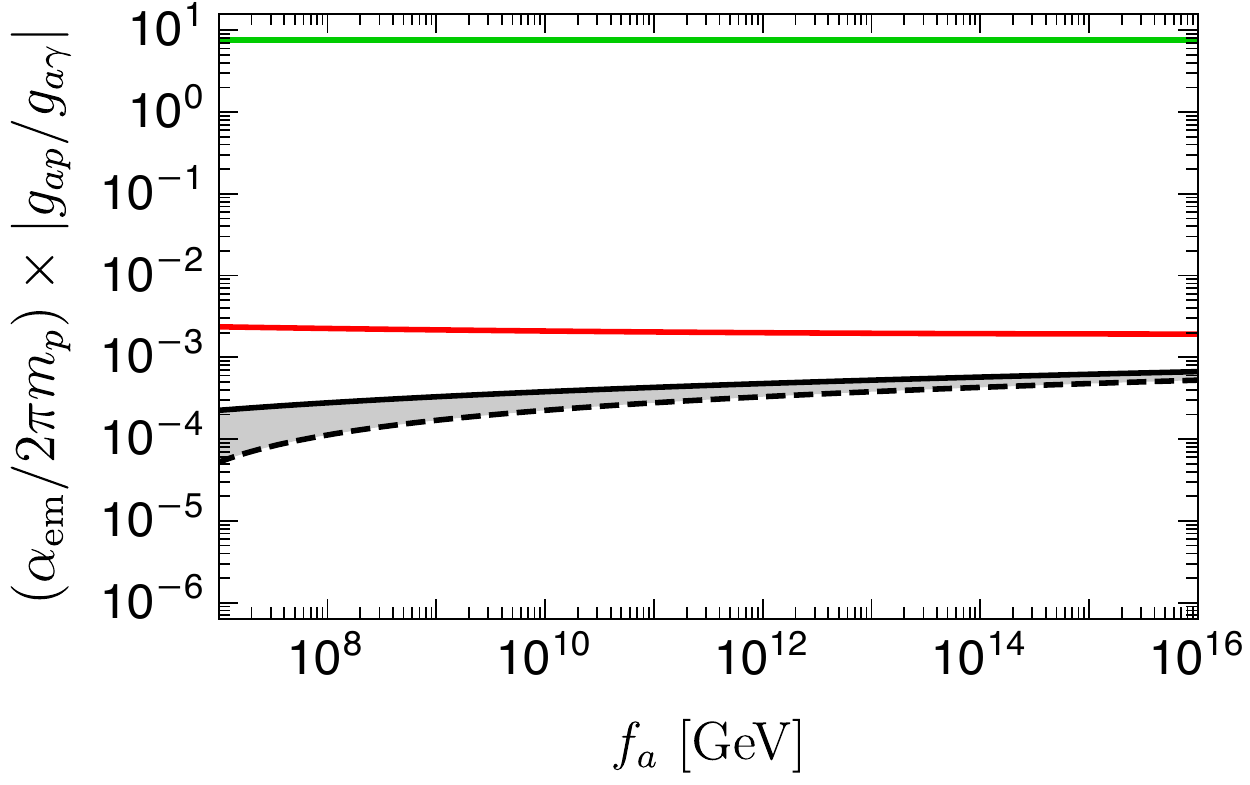} 
      \includegraphics[width=0.5 \textwidth]{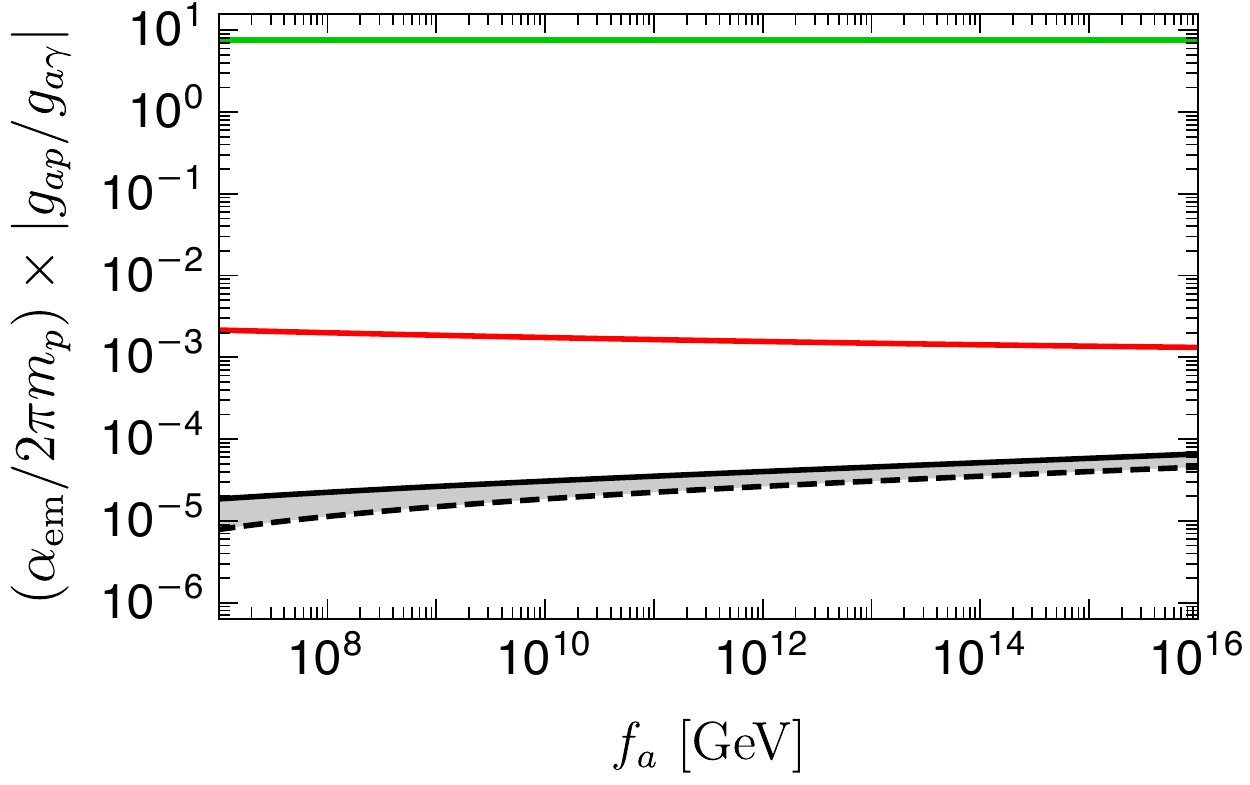} \\
      \includegraphics[width=0.5 \textwidth]{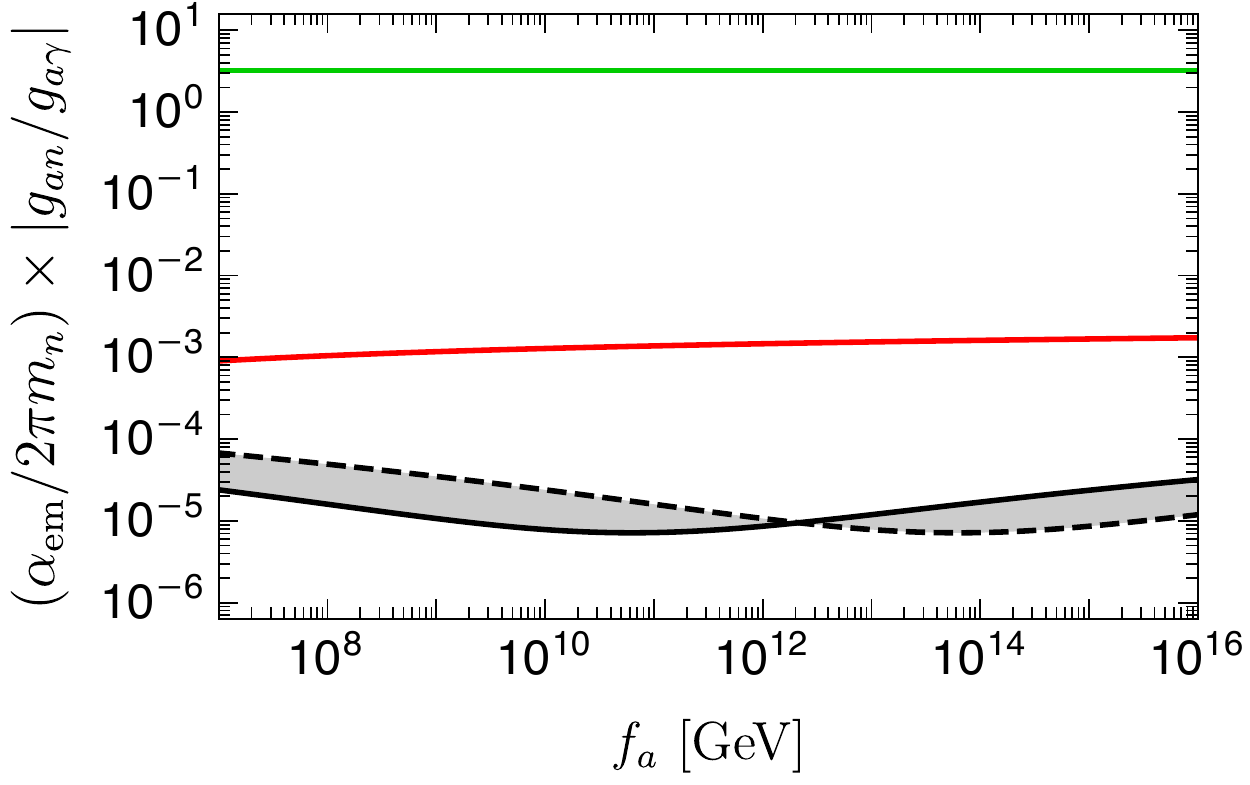} 
      \includegraphics[width=0.5 \textwidth]{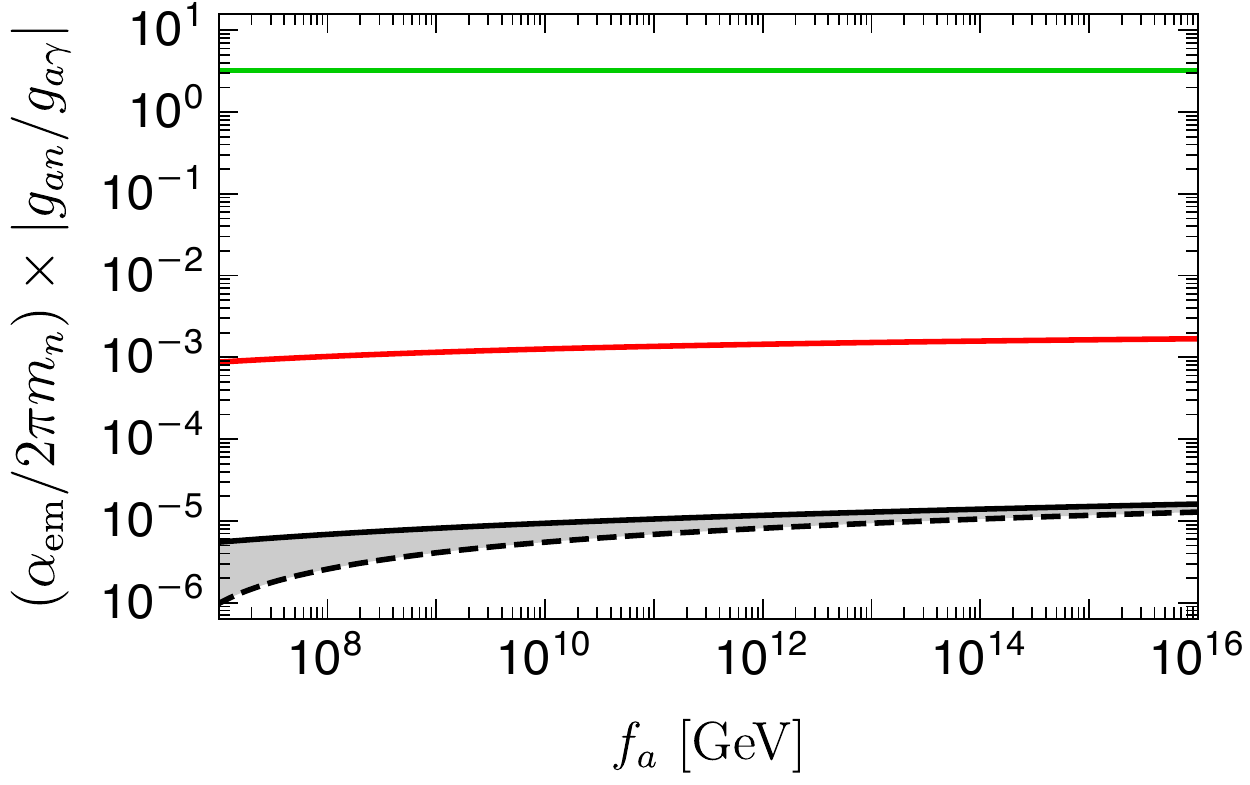} \\
      \includegraphics[width=0.5 \textwidth]{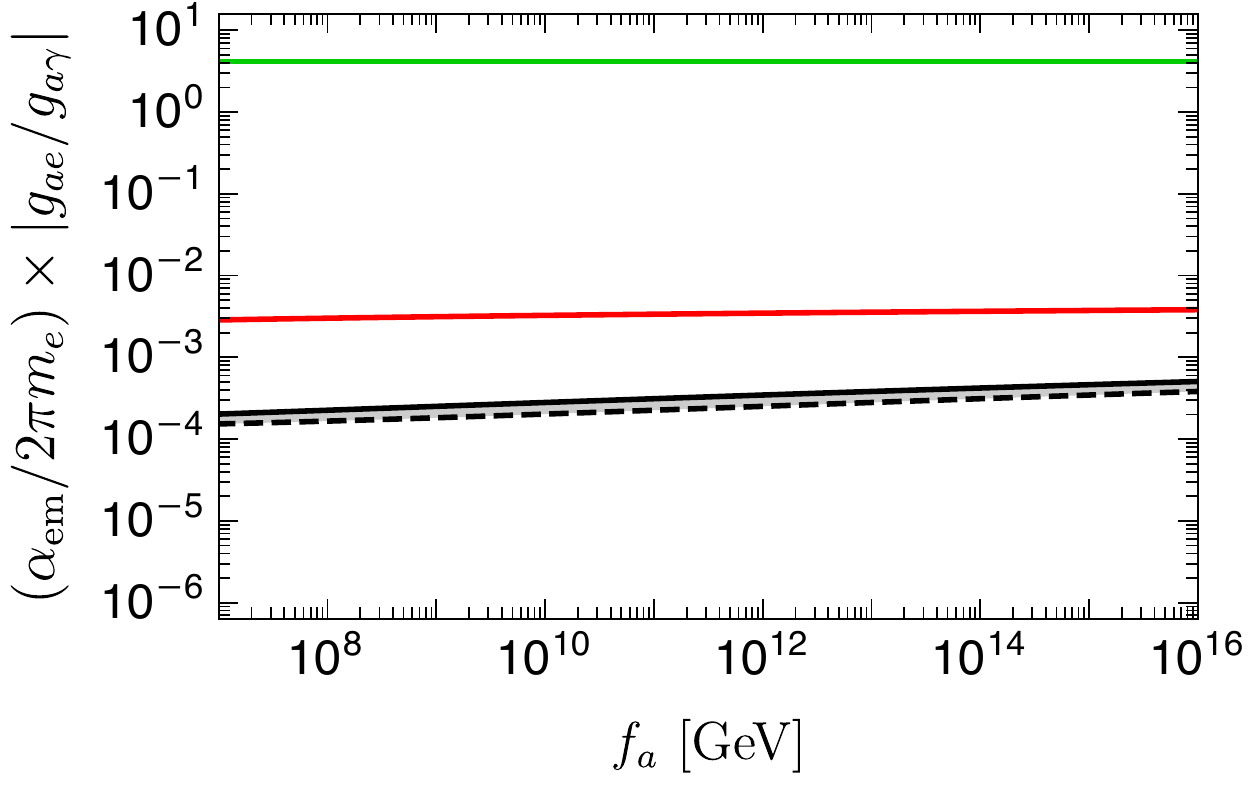} 
      \includegraphics[width=0.5 \textwidth]{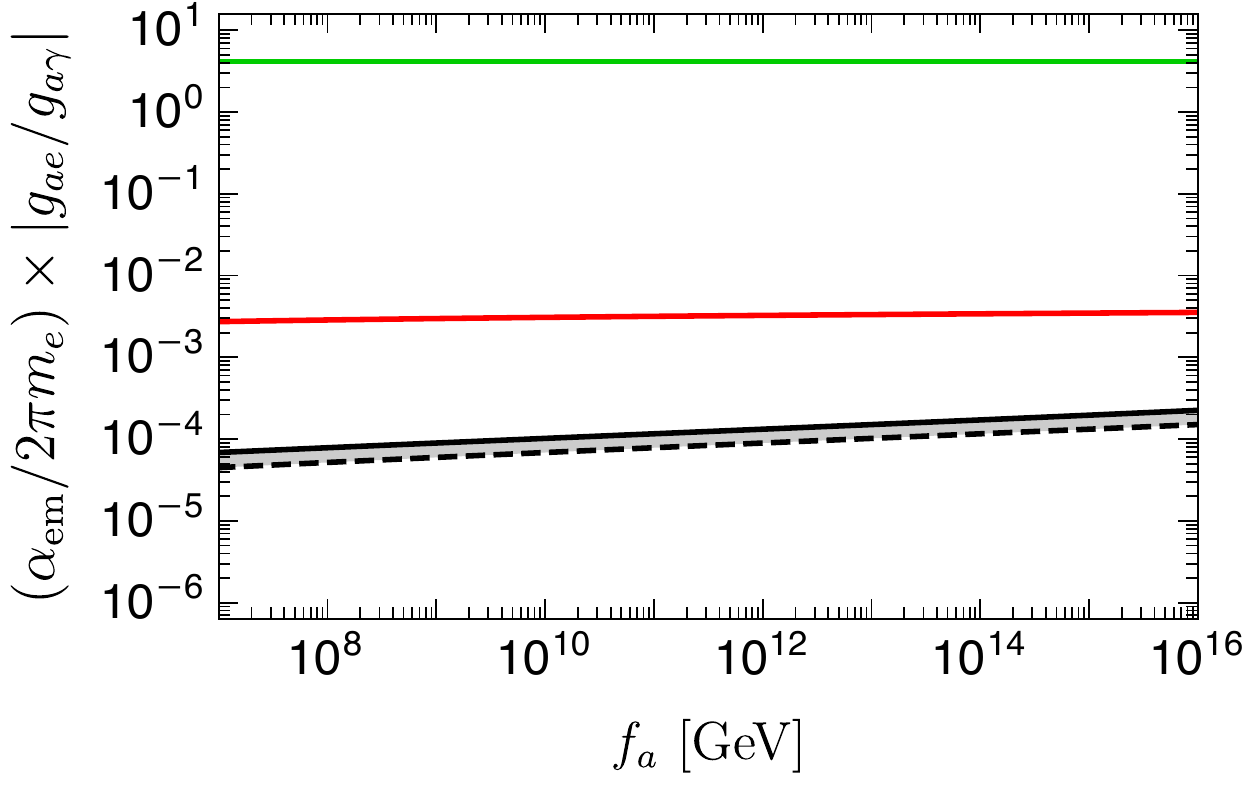} 
  \end{tabular} 
  \caption{ The predicted ratios of the ALP couplings from the minimal axion models with $c_W=1 \,(c_G=c_B=0)$ (left) and $c_B=1 \,(c_G=c_W=0)$ (right) within the MSSM framework (green: DFSZ-like axions, black: KSVZ-like axions, red: string-theoretic axions). For KSVZ-like axions, the heavy color-singlet exotic fermion charged under $SU(2)_L$ (left) or $U(1)_Y$ (right) is assumed to have a mass between $10^{-3} f_a$ (dashed black) and $f_a$ (solid black). 
  For the MSSM parameters, we take $\tan \beta = 10$ and $m_{\rm SUSY}=10$ TeV. }
 \label{ratio_plot} \label{ALPplot}
 \end{figure}

On the other hand, for axions with $c_G = 0$, but $c_W\neq 0$ and/or $c_B\neq 0$, e.g. an ALP coupled to the photon without a coupling to the gluons,  Eqs. (\ref{g_DFSZ})-(\ref{g_string}) tell us that all the three low energy axion couplings may distinguish among the different classes of ALP models.  In Fig. \ref{ALPplot}, we numerically plot the predicted coupling ratios $g_{aY}/g_{a\gamma}\,(Y=p, n, e)$ for each class of ALP. As an illustrative example, we use the same parameter values as in Fig. \ref{QCDaxionplot} but with different $c_A \, (A=G, W, B)$, i.e. $c_G=c_B=0$ and $c_W=1$ for the left panel, while $c_G=c_W=0$ and $c_B=1$ for the right panel. For DFSZ-like ALP, a specific model for such parameter values may include  additional heavy exotic quarks that cancel the gluon anomaly of the model in Eq. (\ref{DFSZtree}) and Eq. (\ref{SUSYDFSZpar}). For KSVZ-like ALP, we use a similar model as in Eq. (\ref{KSVZmod}) with a heavy color-singlet exotic fermion charged under $SU(2)_L$ only (for the left plot) or $U(1)_Y$ only (for the right plot), which is assumed to have a mass between $10^{-3} f_a$ and $f_a$.
Fig. \ref{ALPplot} shows that the axion-proton coupling may be yet hard to discriminate string-theoretic ALPs from KSVZ-like ALPs for $c_W \neq 0$ since the radiative correction becomes sizable if $f_a$ is near the GUT scale.

\section{Conclusions \label{sec:CC}}

Axions are compact scalar fields postulated to solve various issues in particle physics and cosmology including the strong CP problem and dark matter.     
Axions may originate from  
the phase of complex scalar fields (field-theoretic axions) or from  the zero modes of an antisymmetric tensor ($p$-form) gauge field (string-theoretic axions) which couples to
a $(p-1)$-brane in the underlying UV theory. 
Axion couplings which are most relevant for an experimental detection of axions are those to the photon, nucleons and electron
 at scale well below the QCD scale,  i.e. $g_{aX}$ ($X=\gamma, p, n, e$)
which are determined mostly by the quantized couplings to the photon and gluons, as well as the couplings to the light quarks and electron around the QCD scale.
Then, depending on  the UV origin of the field variable and the pattern of 
low energy couplings, axions can be categorized into three  classes: i) field-theoretic DFSZ-like axions with $R_0={\cal O}(1)$,
ii) field-theoretic KSVZ-like axions with $R_0\lesssim {\cal O}((g_{\rm GUT}^2/16\pi^2)^2)$, 
and  iii) string-theoretic axions 
with $R_0={\cal O}(g_{\rm GUT}^2/16\pi^2)$, where $g_{\rm GUT}$ is the gauge coupling at scales around the axion decay constant $f_a$, and $R_0$ denotes the ratio 
(in an appropriate unit) between the coupling 
to the photon (or gluons)
and the {\it tree-level} couplings to the light quarks and electron.  
Compared to the other two, DFSZ-like axions have a clearly distinguishable pattern of $g_{aX}$
($X=\gamma, p, n, e$). However
 it requires a careful analysis of radiative corrections to {the axion couplings to the light quarks and electron 
to discriminate string-theoretic axions from KSVZ-like axions by experimental measurements
of $g_{aX}$ or their ratios.}

 With this motivation, we first performed a generic renormalization group analysis for the axion couplings to matter fermions and Higgs fields
 while taking into account that the Standard Model can be extended to its supersymmetric extension at a scale below $f_a$.
We then applied our results  to string-theoretic axions  and KSVZ-like axions
 to examine if they 
 can have a distinguishable pattern of  $g_{aX}$ ($X=\gamma, p, n, e$).
 We find that discriminating string-theoretic QCD axion  from KSVZ-like QCD axion appears to be quite challenging, but yet it looks feasible with a significantly different value of the coupling ratio $g_{ae}/g_{a\gamma}$. 
 For an axion-like particle (ALP) which does not couple to the gluons, but has a non-zero coupling to the photon, the prospect for discrimination is much more promising: 
all of the three couplings ratios $g_{aY}/g_{a\gamma}$ ($Y=p,n,e$) for 
string-theoretic ALP and KSVZ-like ALP differ by about one order of magnitude.  We also find that the coupling of KSVZ-like QCD axion to the electron is dominated
 by a three-loop contribution involving the heavy exotic quark, gluons, top quarks and Higgs doublet,
 which was not noticed in the previous studies.

\section*{Acknowledgments}
This work was supported by IBS under the project code, IBS-R018-D1.

\bibliography{axion_RGs}
\bibliographystyle{utphys}

\end{document}